\documentclass[11pt,preprint]{aastex} 
\usepackage{natbib}
\usepackage{emulateapj5}

\slugcomment{To Appear in the Ap.J. v.647, no.1 (Aug. 10, 2006)}
\shortauthors{Flohic et~al.}
\shorttitle{19 LINERs Observed With {\it Chandra}}

\begin{document}

\newcounter{species}
\def\ion#1#2{\setcounter{species}{#2}#1$\;${\sc\roman{species}}\relax}

\def\ls{\lower 2pt \hbox{$\;\scriptscriptstyle \buildrel<\over\sim\;$}} 
\def\gs{\lower 2pt \hbox{$\;\scriptscriptstyle \buildrel>\over\sim\;$}} 

\def\kms{\ifmmode{~{\rm km~s^{-1}}}\else{~km~s$^{-1}$}\fi}
\def\ergs{\ifmmode{~{\rm erg~s^{-1}}}\else{~erg~s$^{-1}$}\fi}
\def\Msol{\ifmmode{{\rm M}_{\mathord\odot}}\else{M$_{\mathord\odot}$}\fi}
\def\solar{\ifmmode_{\mathord\odot}\else$_{\mathord\odot}$\fi}

\def\REdd{{\cal R}_{\rm Edd}}


\title{The Central Engines of 19 LINERs as Viewed by \it{Chandra}}

\author{H\'el\`ene M. L. G. Flohic\altaffilmark{1}, Michael
Eracleous\altaffilmark{1}, George Chartas\altaffilmark{1}, Joseph
C. Shields\altaffilmark{2}, \& Edward C. Moran\altaffilmark{3}}

\altaffiltext{1}{Department of Astronomy \& Astrophysics, The
Pennsylvania State University, 525 Davey Lab, University Park, PA
16802 e-mail: {\tt flohic, mce, chartas @astro.psu.edu}}

\altaffiltext{2}{Department of Physics \& Astronomy, Ohio University,
251B Clippinger Labs, Athens OH, 45701-2979 e-mail: {\tt
shields@phy.ohiou.edu}}

\altaffiltext{3}{Astronomy Department, Wesleyan University,
Middletown, CT 06459 e-mail: {\tt ecm@astro.wesleyean.edu}}
 
\begin{abstract}
Using archival {\it Chandra} observations of 19 LINERs we explore the
X-ray properties of their inner kiloparsec to determine the origin of
their nuclear X-ray emission, to investigate the presence of an AGN,
and to identify the power source of the optical emission lines .  The
relative numbers of LINER types in our sample are similar to those in
optical spectroscopic surveys. We find that diffuse, thermal emission
is very common and is concentrated within the central few hundred
parsec. The average spectra of the hot gas in spirals and ellipticals
are very similar to those of normal galaxies. They can be fitted with
a thermal plasma ($kT\sim 0.5$~keV) plus a power law (photon index of
1.3-1.5) model. There are on average 3 detected point sources in their
inner kiloparsec with $10^{37}<L_{0.5-10\;\rm{keV}}<10^{40}$ erg
s$^{-1}$. The average cumulative luminosity functions for sources in
spirals and ellipticals are identical to those of normal galaxies. In
the innermost circle of 2{\farcs}5 radius in each galaxy we find an
AGN in 12 of the 19 galaxies.  The AGNs contribute a median of 60\% of
the 0.5-10 keV luminosity of the central 2{\farcs}5 region, they have
luminosities of $10^{37}$--$10^{39}~\ergs$ (Eddington ratios of
$10^{-8}$ to $10^{-5}$). The ionizing luminosity of the AGNs is not
enough to power the observed optical emission lines in this particular
sample. Thus, we suggest that the lines are powered either by the
mechanical interaction of an AGN jet (or wind) with the circumnuclear
gas, or by stellar processes, such as photoionization by post-AGB
stars from an old or intermediate-age population (in most cases) or by
young stars (in a few cases).

\end{abstract}


\section{Introduction} 

In 1980, Heckman identified a class of galactic nuclei whose optical
emission-line properties differ from those of classical active
galactic nuclei (AGN) and \ion{H}{2} regions. These objects have narrow
emission lines whose relative intensities are indicative of a
relatively low ionization state of the emitting gas. Heckman (1980)
named these objects low-ionization nuclear emission-line regions
(LINERs) and defined them using the ratios of the optical forbidden
lines of oxygen:
[\ion{O}{2}]$\lambda$3727/[\ion{O}{3}]$\lambda$5007$>$1 and
[\ion{O}{1}]$\lambda$6300/[\ion{O}{3}]$\lambda$5007$>$0.33. Since the
division is observationally motivated, any object satisfying one of
the criteria and close to satisfying the other one is often called a
LINER. Recognizing that this classification is somewhat arbitrary,
Heckman (1980) noted the existence of ``transition'' objects whose
spectra are intermediate between those of pure LINERs and \ion{H}{2}
regions.
 
The original idea proposed by Heckman (1980) to explain LINERs was
that the emission lines are produced by shocks in the nucleus of the
host galaxy. In fact, the optical spectrum arising from a shock-heated
gas closely resembles the spectrum of LINERs \citep{DS95,
Dal96}. Similar emission could be produced by turbulent mixing layers
between cold and hot ISM components \citep{F96}. Another possibility
is that the lines originate in dense gas photo-ionized by X-rays
(Ferland \& Netzer 1983; Halpern \& Steiner 1983). The ionizing source
can be a low-luminosity AGN (LLAGN), an idea supported by the
discovery of broad emission lines (FWHM$>10^3$ km s$^{-1}$) in the
spectra of many LINERs (e.g., Filippenko \& Halpern 1984; Ho et
al. 1997b; Stocrchi-Bergmann, Baldwin, \& Wilson 1993; Bower et
al. 1996; Barth et al. 1999; Shields et al.  2000; Barth et al. 2001;
Eracleous \& Halpern 2001). Alternatively, the source of ionizing
photons could be the Wolf-Rayet or OB stars in a compact, young star
cluster in the nucleus of the host galaxy (Terlevich \& Melnick 1985;
Filippenko \& Terlevich 1992; Shields 1992; Barth \& Shields
2000). Yet another possibility is that post-AGB stars from an old
stellar population provide the ionizing photons that power the
LINER-like emission lines (Binette et~al. 1994).
 
Recent spectroscopic surveys have shown that LINERs are very common in
local galaxies, found in at least 30\% of all galaxies and 50\% of
early-type spirals (see review by Ho 1996, and references
therein). Since LINERs are so common, determining the nature of their
power source is important and could have far-reaching implications. If
most LINERs are genuine AGNs, they would be the most common type of
AGN in the local universe and would represent the faint end of the AGN
luminosity function. The study of LINERs could shed light on the
evolution of AGN and their host galaxies since they are likely to
represent the end state of quasar evolution. It would also set a lower
limit to the frequency of accreting supermassive black holes (SMBHs)
in galaxies. Moreover, the spectral energy distribution of LINERs is
different from that of classical AGNs implying that, if they are AGNs,
their mode of accretion might also be different \citep{Lal96,
H99}. Suggested accretion modes are advection dominated accretion
flows (ADAFs; Narayan \& Yi 1994, 1995), convection-dominated
accretion Flows (CDAF; Quataert \& Narayan 1999) and adiabatic
inflow-outflow solutions (ADIOS; Blandford, \& Begelman 1999). If
LINERs are powered by hot stars or stellar remnants, one can ask why
their spectra are different from disk \ion{H}{2} regions. Moreover,
the presence of a central star cluster influences the chemical and
dynamical evolution of the bulge of the galaxy \citep{B00, C99,
Combes01}, so the frequency of compact nuclear star clusters is of
significant interest.

In view of the importance of determining the nature of the central
engines of LINERs, we have undertaken the analysis of a large sample
of archival {\it Chandra} observations of galaxies hosting LINERs. The
X-ray morphology of the central kiloparsec of the nucleus at the
angular resolution of {\it Chandra} can give a first hint to the power
source. An unresolved hard X-ray source at the center of the galaxy is
a good indication of an AGN, while a cluster of point sources and
knotty diffuse emission suggests stellar phenomena. The high spatial
resolution of {\it Chandra} and its spectroscopic capabilities also
make it possible to extract the spectra of the spatially-distinct
components (point sources and diffuse emission) to confirm their
nature and to evaluate their contribution to the energy budget. This
is an important issue because some LINERs may be powered by a
combination of an AGN and a starburst (hereafter ``composites''; e.g.,
Veron et~al. 1981, Moran et~al. 1996, Fernandes et~al. 2004).

Previous surveys with {\it Chandra} have studied the X-ray properties
of LINERs with different techniques, purposes or samples than our
study. Terashima \& Wilson (2003) used snapshot observations of LINERs
with high radio fluxes in order to search for AGNs and determine their
properties. Because of their short exposure times, they were able to
detect AGNs with X-ray luminosities greater than $10^{38} -
10^{39} \ergs$.  Satyapal et al. (2005) and Dudik et al. (2005) used
a combination of long and short {\it Chandra} observations to assess
the presence of an AGN in LINERs and IRAS observations to determine
the presence of star formation but they do not examine the energy
bugdet of these sources.  They did, however, determine that at least
50\% of LINERs harbor a central hard X-ray source consistent with an
AGN. Filho et al. (2004) find that a similar fraction of LINERs harbor
AGNs by means of radio observations of 16 LINERs and {\it Chandra}
observations of 8 of them. They focused on the properties of the
nucleus and argued that LINERs are powered by an AGN, which is often
undetected in the X-rays because the flux is highly
absorbed. Eracleous et al. (2002) studied three LINERs addressing the
same questions that we address here. Out of those three LINERs, one
was powered by an AGN alone and two by stellar processes alone
suggesting a diversity in the nature of the ionization sources of
LINERs. In this paper we continue and extend the work of Eracleous et
al. (2002). Some of the objects in our sample were studied
individually using the same {\it Chandra} data that we use here.
However, we repeat the analysis for the sake of uniformity and we add
two important elements in our study: a detailed examination of the
circumnuclear regions, and a careful consideration of the energy
budget. The objects in our sample are unlike some of the notorious
LINERs whose nature is clear-cut, but they are typical of the greater
population of LINERs.

In this paper we present the results and conclusions of our study,
organized as follows.  The sample and initial data reduction
procedures are described in \S\ref{sec:targets}. Model fits to the
spectra of individual discrete sources and the average spectra of
diffuse sources are described in \S\ref{sec:analysis}. In
\S\ref{sec:populations} we analyze the properties of the point source
populations collectively and in \S\ref{sec:individual}, we evaluate
the power source of each galaxy from the sample individually.
Finally, we summarize the results and discuss the implications in
\S\ref{sec:discussion}.

\section{Targets and Data Screening\label{sec:targets}} 

We examined observations of LINERs in the {\it Chandra} archive up to
2002 December and selected a sample of 19 galaxies hosting a pure
LINER or a transition object (Ho et~al. 1997a) based on their distance
($< 25$ Mpc) and their exposure time ($> 15$ ks). These selection
criteria ensure good spatial resolution in the central kiloparsec of
each galaxy, a high signal-to-noise ratio (S/N) and a low detection
limit (the limiting 0.5--10~keV point source luminosity is in the
range $3.5\times 10^{36}$ to $3.4\times 10^{37}\ergs$). Most (14 out
of 19) of the galaxies were observed for reasons unrelated to the
presence of a LINER, thus the sample is largely unbiased with respect
to the LINER properties. In \S\ref{sec:discussion}, we use the
distribution of LINER types in our sample, to argue that it is
representative of the general population of LINERs.

The properties of the host galaxies and the exposure times are
summarized in Table~1. The LINER type follows the classification of Ho
et~al. (1997a) where L1 and L2 are ``pure'' LINERs. L1 have fairly
strong narrow emission lines combined with weak broad H$\alpha$ but no
broad H$\beta$ while L2 have only narrow emission
lines. ``Transition'' objects with line ratios intermediate between
those of \ion{H}{2} regions and LINERs are referred to as T2.  The
distance is taken from Ho et~al. (1997a) where they assume
$H_{0}=75$~km~s$^{-1}$~Mpc$^{-1}$. Column~5 gives the Galactic
hydrogen column density ($N_{\rm H}$) in the direction of the center
of each galaxy \citep{DL90}. NGC~1553 was not included in the Ho et
al. (1997a) sample, but was listed as a LINER by Phillips et
al. (1986). Among the spiral galaxies, NGC~2681, NGC~3507, NGC~4314,
NGC~5055 and NGC~7331 are viewed approximately face-on while NGC~4111
is approximately edge-on.

Table~1 also includes the H$\alpha$ luminosities of the target
galaxies, as well as the estimated black hole masses. The H$\alpha$
luminosities were taken from from Ho et al. (1997a), who measured them
from spectra obtained through a $2^{\prime\prime}\times
4^{\prime\prime}$ aperture. NGC 1553 is the only exception; its
H$\alpha$ luminosity was taken from Phillips et al. (1986). The black
hole masses were estimated from the stellar velocity
dispersion\footnote{Stellar velocity dispersions were taken from the
Hypercat database, available electronically at {\tt
http://www-obs.univ-lyon1.fr/hypercat.}} using the prescription of
Tremaine et al. (2002). The derived black hole masses are between
$10^7$ and $6.3\times 10^8~\Msol$, with an average of $1.2\times
10^8~\Msol$ and typical uncertainties of a factor of 2 (or $\pm 0.3$
dex; see discussion in Barth et al. 2002).
 
All the target galaxies were observed with the Advanced CCD Imaging
Spectrometer (ACIS; Garmire et~al. 2003). The nuclei of most galaxies
were located near the S3 CCD aimpoint, which did not suffer
significant radiation damage during the early stages of the
mission. Only NGC~3607 and NGC~3608 were on a different CCD: I2 and I0
respectively. The initial data screening was carried out with the CIAO
software package v 3.2.1\footnote{{\it Chandra} Interactive Analysis
of Observations (CIAO), {\tt http://cxc.harvard.edu/ciao/}}. It
consisted of the selection of events with grades 0, 2, 3, 4 and 6 and
the exclusion of events occurring during times of poor aspect solution
and background flares. We also removed the 0\farcs 5 pixel
randomization and checked for aspect-solution errors. We also applied
a CTI correction to the data for the two ACIS-I sources. We corrected
for pointing offsets using the appropriate information on the {\it
Chandra} X-ray Center's web-site in order to improve the astrometry
\footnote{\tt
http://cxc.harvard.edu/cal/ASPECT/fix\_offset/fix\_offset.cgi}.

After screening, we produced an event file of the region centered on
the nucleus with a 1~kpc radius. The coordinates of the stellar
nucleus were determined from isophotal contours of infrared images
obtained with 2MASS. When coordinates derived from radio measurements
or dynamical measurements were available, they were compared with the
2MASS coordinates and we found that they were always in agreement
within errors (except for the VLBA coordinates of NGC~4552, which were
1.25 $\sigma$ away from the 2MASS coordinates). For reference we note
that the typical astrometric uncertainty from 2MASS isophotal fitting
is 1{\farcs}25, while the typical astrometric uncertainties from radio
observations are 25~mas (VLA), 5~mas (VLBA), and 1.3~mas (VLBI).  In
comparison, the astrometric uncertainty of {\it Chandra} observations
is 0{\farcs}5. The nucleus of all the galaxies is at most 7\arcsec\
away from the ACIS-S aimpoint, except for NGC~3607 and NGC~3608, whose
nuclei are respectively 211\arcsec\ and 140\arcsec\ from the ACIS-I
aimpoint.

\section{Data Analysis\label{sec:analysis}}

We begin the analysis of the data by constructing images of the
central region of each galaxy, which we present in
\S\ref{sec:images}, below.  This is followed by fitting models to the
spectra of bright, discrete sources in the nuclei of the host galaxies
(discussed in \S\ref{sec:fits}) and to the average spectra of diffuse
emission in spiral and elliptical galaxies (presented in
\S\ref{sec:diffuse}).

\subsection{Images and Morphology\label{sec:images}}

From the processed event files, we produced images of a region of
radius 1~kpc around the center of each galaxy (hereafter, the ``central
kiloparsec'') in two different bands: 0.5--2~keV and 2--10~keV. We also
produced hardness ratio maps by combining the soft ($S$) and hard
($H$) images in the following way: $HR=(H-S)/(H+S)$. For pixels which
had no counts in either the soft or the hard band, we adopted a
hardness ratio equal to the average of the image.

Figure~1 shows the image of the central kiloparsec of each target
galaxy in the soft band (0.5--2 keV) and the hard band (2--10 keV) and
the hardness ratio map. The morphology of each galaxy is summarized in
Table~2. The first column gives the name of the galaxy, while the
second column gives the number of point sources that were identified
by the {\tt wavdetect} algorithm \citep{Freemanal02} in the 0.5--10 keV band. To prevent {\tt
wavdetect} from confusing fluctuations in diffuse emission with true
discrete sources, we ran the algorithm on a $4\times4$~kpc region with
a threshold of 10$^{-6}$.  An S in the third column indicates that a
point source is exactly at the center of the galaxy in the 0.5--2~keV
energy band, while an H indicates that a point source in the 2--10~keV
energy band is at the galaxy center. The combination of an S and an
H indicates that the central source is detected in both energy
bands. To determine if a source is coincident with the center of a
galaxy, we checked whether the {\it Chandra} source coordinates
matched the nuclear coordinates (from 2MASS or radio observations)
within the 1-$\sigma$ uncertainty region. The fourth column gives the
same information as the third column but for the diffuse emission.

The morphology of the nuclear region as a function of energy can give a
first hint for the power source of the LINERs. A hard central source
suggests an AGN and a cluster of point sources or a knotty diffuse
emission suggests stellar processes. When both are observed at the
same time, this hints at a combination of a stellar processes and an
AGN.

\subsection{Fits to Spectra of Discrete Sources\label{sec:fits}}

We extracted the spectra of all the point sources using the {\tt ACIS
Extract} software \citep[{\tt AE},][]{AE}. {\tt AE} uses polygonal
extraction regions which approximate contours of the {\it Chandra} -ACIS
point spread function (PSF). A PSF contour is selected such that the
fraction of the PSF energy enclosed is approximately 95\%. The
background was measured as close to the point sources as possible in
order to decrease the contamination from diffuse emission, which
is fairly common.  Spectral fitting was carried out using
XSPEC~v.11.2.0x \citep{A96}. The spectra were truncated at 0.5~keV at
the low-energy end and between 2 and 10~keV, depending on the S/N, at
the high-energy end. The spectral models were also modified by
interstellar photoelectric absorption with cross sections as given by
Morrison \& McCammon (1983).

Spectra of faint sources with 20--200 counts were binned adaptively by
{\tt AE}. These spectra were then fitted with an absorbed power-law
model in order to determine the point source flux. Sources with less
than 20 counts could not be fitted because of the small number of
degrees of freedom. For these sources the luminosity was estimated by
scaling a power law model ($\Gamma = 1.8$, close to the median
$\Gamma$ for the individual fits to the point sources) absorbed by the
average Galactic column density ($N_{\rm H}=2\times 10^{20}~{\rm
cm}^{-2}$) so that the predicted count rate equals the observed count
rate. We chose this nominal value of $N_{\rm H}$ because it is an
average Galactic column for the sample and a small variation in the
$N_{\rm H}$ ($<2\times 10^{19}~{\rm cm}^{-2}$) did not have an impact
on the resulting X-ray flux. Using this model, we also estimated the
lowest detectable 0.5--10~keV point source luminosity for each galaxy,
which we list in Table~2. The luminosities of all the discrete sources
are used in our study of the luminosity functions in
\S\ref{sec:populations}.

We estimated the uncertainties in our {\it faint} ($< 200$ counts) X-ray point source
luminosities by varying $\Gamma$ by $\pm 0.2$ for a power-law absorbed
by the Galactic column. The 0.5--10 keV luminosity then varied by 5--10\%. When
varying $\Gamma$ by $\pm 0.7$, the luminosity changes by
30\%--40\%. When increasing the absorbing column density by one order
of magnitude, the luminosities increased by
5--25\%. Increasing the absorbing column density by two orders of
magnitude dramatically increased the luminosity by 100\%--200\%. A
small variation in the absorbing column density (20\%) did not have an
impact on the X-ray luminosity estimate. Thus, based on the above
estimates, and assuming that an absorbed power-law model provides an
acceptable description of the spectrum, the typical uncertainty in the
X-ray luminosities we derive is \ls 30\%, unless the
absorbing column density is grossly underestimated.

The spectra of sources with more than 200 counts were binned with a
minimum of 20 counts per bin and different models were fitted to the
spectra in order to determine the nature of the source and the
luminosity. There is never more than one source per galaxy with more
than 200 counts and it is usually the central point source.  It is
important to find the most apropriate model for the X-ray spectrum so
as to determine the absorbing column reliably. Systematic errors in
the absorbing column can lead to large systematic errors in the source
luminosity. Thus, we have tested the following models:

\begin{itemize}

\item A blackbody spectrum ({\tt bb}), with the temperature ($kT$) and
the normalization as free parameters.

\item A comptonized blackbody spectrum ({\tt compbb}, after Nishimura
et~al. 1986) with the seed blackbody photon temperature ($kT_{\rm
seed}$), the electron temperature of the hot plasma ($kT_{\rm e}$),
the optical depth of the plasma ($\tau$), and the normalization as
free parameters.

\item A spectrum from an accretion disk consisting of multiple
blackbody components ({\tt diskbb}, after Mitsuda et~al. 1984;
Makishima et al. 1986) with the temperature at the inner disk radius
($kT_{\rm in}$) and the normalization as free parameters.

\item An emission spectrum from hot diffuse gas with up to three
components at different temperatures ({\tt mekal}, {\tt mekal2T}, {\tt
mekal3T}; after Mewe et~al. 1985, 1986; Kaastra 1992; Liedhal
et~al. 1995) with the plasma temperature ($kT_i,~i=1,2,3$), the
metallicity ($Z$, initially set to solar), and the normalization as
free parameters. The redshift was set to 0 since the galaxies are all
nearby.

\item A power-law spectrum ({\tt pow}) with the photon index of the
power-law ($\Gamma$), and the normalization as free parameters.

\item A spectrum from an accretion disk around a black hole in an
X-ray binary Compton scattered from in a converging, plasma
flow ({\tt bmc}, after Titarchuk et~al. 1997; Titarchuk \& Zannias
1998; Laurent \& Titarchuk 1999; Borozdin et~al. 1999; Schrader \&
Titarchuk 1999) with the temperature of the thermal photon source
($kT_{\rm seed}$), the index of the electron power-law energy
distribution ($\alpha$), the illumination parameter, and the
normalization as free parameters.

\end{itemize}

We have also tested certain linear combinations of these models such
as blackbody plus power law or {\tt mekal} plus power law (as well as
multiple {\tt mekal} components at different temperatures).

These models can describe the spectra of X-ray binaries, accreting
supermassive black holes or thermal diffuse emission depending on
their luminosity and fit parameters. For low mass X-ray binaries, the shape of the spectrum depends on the spectral state
\citep{Barret01}. In the high soft state, the spectrum can be modeled
as a soft component ({\tt bb} with $kT< 1$~keV or {\tt diskbb} with
$kT_{\rm in}$ between 0.5 and 1.5 keV) and a comptonized component
({\tt compbb}) with $\tau\sim 5 - 15$, $kT_{\rm e}\sim 3$~keV and
$kT_{\rm seed}$ between 0.3 and 1.5 keV and with $L_{\rm 0.5-10\;
keV}$ between $10^{37}$ and $10^{38}\ergs$. In the low hard hard
state, the spectra low-mass X-ray binaries can be modeled as a soft
component and a power-law tail but with with lower relative
luminosities (and absolute luminosities below the detection limit in
our sample galaxies). High-mass X-ray binaries have a hard spectrum
with $L_{\rm 0.5-10 \; keV} < 10^{39}\ergs$ that can be modeled as a
power-law of $\Gamma\sim 1.2$ \citep{PR02}. Accreting supermassive
black hole such as those found at the center of Seyfert galaxies are
observed to have a power-law spectrum with $\Gamma\sim 2$. Their
luminosities are typically greater than $10^{39}$ erg s$^{-1}$ (see
Mushotzky, Done, \& Pounds, 1993, and references therein), but less
luminous AGNs are well known to exist, especially in LINERs.

Table~3 summarizes the two best-fitting models for the spectrum of the
central object in each galaxy, or the brightest off-nuclear
object. Fits with an acceptable value of the reduced $\chi^{2}$ but
with unphysical parameters ($N_{\rm H}$ lower than that listed in
Table~1, unreasonable photon index or temperature) were not included
in Table 3. Column~2 of Table 3 gives the models that provided the
best two fits and the values of the model parameters. Column~3 gives
the $\chi^{2}$ and the number of degrees of freedom, and column~4 the
luminosity calculated for the model in the 0.5--10~keV range. For all
these models, we assumed a solar metal abundance. In
Figure~\ref{fig:spectra}, we show the best-fitting model for each
spectrum (the first one listed for each object in Table 3) superposed
on the data along with the post-fit residuals. When there were
significant residuals, we tried to modify the metallicity using the
values given by Terlevich \& Forbes (2002) and letting it vary, but it
never suppressed the residuals, nor did it improve the reduced
$\chi^{2}$, and was not supported by a F-test; therefore we kept solar
abundances. We explored the effects of assuming Solar abundance
in objects with sub-solar or super-solar abundances by increasing and
lowering the abundace and measuring the change in the other fit
parameters. We found that the other fit parameters (column density,
power-law index, temperature, luminosity) varied very slightly ($<
5\%$) when the abundance increased or decreased by one order of
magnitude.

\subsection{Diffuse Emission\label{sec:diffuse}}

In 70\% of the galaxies of our sample we find diffuse emission in the
central kiloparsec. The frequent presence of diffuse emission was
already noted by Halderson et al. (2001). In 87\% of the galaxies that
have diffuse emission, this emission contributes over 50\% of the
observed X-ray luminosity.  The flux from the diffuse emission in each
galaxy was not high enough to yield a high S/N spectrum individually
except for NGC~3628. Thus, we stacked the spectra from all the
galaxies. First, we extracted the spectrum from a 1 kpc radius around
the center of each galaxy by carefully subtracting all the point
sources and by choosing a background region free of diffuse emission
and point sources. In the process of subtracting the point sources we
also subtracted some diffuse emission, which makes our estimate of the
diffuse emission luminosity a lower limit, strictly speaking. Since
the number of point sources in each galaxy is small (3 on average), we
estimate that this lower limit is not too far from the true
luminosity. Before combining the spectra, we normalized them to a
fiducial total count rate of 0.1 s$^{-1}$ so that the result is a good
representation of the average spectrum of this type of source. Then we
stacked the spectra using the ftool {\tt addspec} and propagated the
errors appropriately. NGC~3607 and NGC~3608 were not included in the
stacking procedure because they were observed with a different CCD and
{\tt addspec} requires that all the observations are made with the
same CCD. We stacked separately the diffuse emission from spiral and
elliptical galaxies. Figure 2 shows the resulting spectra.

The diffuse emission spectra were not well fitted by a two-temperature
component plasma, instead they required a plasma model plus a
power-law, with parameters as given in Table~3. The spectra
themselves, with models superposed are shown in
Figure~\ref{fig:stackedspec}.  The model parameters are very similar
between elliptical and spiral galaxies, $N_{\rm H}\approx (7-9) \times
10^{19}~{\rm cm}^{-2}$, $kt\approx 0.6$~keV, $\Gamma=$1.3--1.5, with solar abundance. We also let the abundance vary as a free
parameter, but the improvement in reduced $\chi^2$ was not significant
according to the F-test. In \S\ref{sec:populations} we estimate the
average luminosity of unresolved X-ray binaries in the inner
kiloparsec and we show that it is comparable to the luminosity of the
power-law spectral component.  We also fitted the diffuse emission
spectrum of NGC~3628 independently and we fount that the two best
models were either a two-temperature plasma model or a power law with
parameters as given in Table~3 (the spectrum is shown in
Fig.~\ref{fig:spectra}). The spectral parameters for this particular
galaxy are in general agreement with what we find for the stacked
spectra, above. We also note that David et al. (2005) studied the
diffuse emission spectrum of one of the galaxies in our sample,
NGC~3379, and found that it could be fitted with a similar model as
our stacked spectra, above.
 
The plasma component of the diffuse emission has a similar temperature
in both types of galaxies with a value that is slightly lower than
what is observed in normal galaxies; $kT=0.6$--0.7 keV for spirals
\citep{Wang01, Wang03, Immler03} and $kT=1$~keV for ellipticals
\citep{Forman85, Trinchieri86, Sal04}.  From the normalization
constant of the XSPEC fit (after rescaling it to the average {\it
observed} count rate), we can calculate the emission measure and then
the density. We find that $n=6.2\times 10^{-2}~{\rm cm}^{-3}$ for the
elliptical galaxies and $n=4.8\times 10^{-2}~{\rm cm}^{-3}$ for the
spiral galaxies. These densities are comparable to those found in
normal galaxies \citep[e.g.,][]{Sal03,Soria06a}. From the stacked
spectra, we can also deduce that the average X-ray luminosity of the
diffuse emission (integrated over the inner kiloparsec) is $\langle
L_{\rm 0.5-10\; keV}\rangle =5.3 \times 10^{38}~{\rm erg~s}^{-1}$ for
the ellipticals and $\langle L_{\rm 0.5-10\; keV}\rangle =8.5\times
10^{38}~{\rm erg~s}^{-1}$ for the spirals. Within uncertainties, the
luminosities of the diffuse emission in spiral and elliptical galaxies
are comparable to each other. We calculated the luminosity of the
diffuse emission for each galaxy individually by scaling the model
luminosity to the count rate for each galaxy. Table~4 summarizes
the 0.5--2 and 2--10 keV luminosity of the diffuse emission within the
central kiloparsec of each galaxy (Columns 2 and 3) and the 0.5--2 and
2--10 keV luminosity of the diffuse emission in the central 2\farcs 5
region (Columns 4 and 5). The last quantity was computed by
integrating the model spectrum after scaling it to the count rate of
each individual; we use this quantity in our evaluation of the power
sources of individual galaxies in \S\ref{sec:individual}.

The sound speed of the hot gas is $c_{\rm s}=(\gamma kT/\mu m_{\rm
H})^{1/2}\approx 250$ km s$^{-1}$ (assuming solar composition, full
ionization and $kT=0.5$ keV; $\gamma$ is the ratio of specific heats
and $\mu$ is the mean molecular weight), slightly larger than the
stellar velocity dispersions in these galaxies.
%
%
The dynamical and cooling time scales of the hot gas are $\tau_{\rm
dyn} = {R/c_s} \approx 10^6$~years and $\tau_{\rm cool} = {3\over
2}nkTV/L_{\rm X}\approx 4\times 10^7$~years, respectively
(assuming that the gas is contained in a sphere of $R=200$~pc). In
view of these time scales and the fact that the sound speed is not
considerably higher than the stellar velocity dispersion, it appears
likely that the gas is bound to the host galaxy and its temperature is
sustained by interaction with the stars. The gas may, in fact,
originate in stellar mass loss. As David et al. (2005) note, the mass
of hot gas in NGC~3379 is approximately $5\times 10^{5}~\Msol$ and it
could have been supplied by stellar mass loss over a period of only
$10^7$~years.

\section{Point Source Population\label{sec:populations}}
 
The LINERs in our sample have between zero and 12 point sources in
their central kiloparsec (including candidate AGNs) with an average
number of approximately 3 point sources and a most frequent value of
two point sources. The point source detection limit for our sample
varies between $3.5\times10^{36}$ and $3.4\times10^{37} {\rm
~erg~s^{-1}}$ in the 0.5--10 keV band (see Table~2). The 0.5--10~keV
luminosity of the detected point sources varies between $1\times
10^{37}$ and $1\times 10^{40}{\rm ~erg~s^{-1}}$ with an average of
$3.3\times 10^{38}{\rm ~erg~s}^{-1}$.

In order to determine whether the point source populations of the
central kiloparsec of LINERs differ from those of normal galaxies, we
constructed luminosity functions (including point sources from all the
galaxies in the sample but excluding candidate AGNs) and compared them
to those of normal galaxies. The normal galaxy comparison sample 
comprises 6 galaxies with a wide range of Hubble types that have weak
optical emission lines and have been observed intensively in the X-ray
band \citep[NGC~1637, NGC~3556, NGC~4365, NGC~4382, NGC~4631 and
NGC~4697;][]{Sal03,Sarazinal01,Wang01,Immler03}.

We constructed the average luminosity functions for the elliptical
galaxies and spiral galaxies separately using point source luminosities
measured as described in \S\ref{sec:fits}. We excluded candidate AGNs
and truncated each luminosity function at the highest detection limit
of galaxies in that category ($6.6\times 10^{37}$ and $3.4\times
10^{37}\ergs$ for ellipticals and spiral respectively). There are 21
sources in the elliptical galaxy luminosity function and seven sources
in the spiral galaxy luminosity function. The resulting cumulative
luminosity functions are shown in Figure 4. We fitted a simple
power-law through the cumulative luminosity function of the sources in
spiral galaxies, which we derived from the differential luminosity
function as follows:
$$
{dN\over dL} \equiv N(L)\equiv {N_0\over L_0}\;
\left({L\over L_0}\right)^{-\alpha} 
$$
\begin{equation}
\quad \Rightarrow \quad
N(>L)\equiv \int_L^{\infty} N(L)\; dL = 
N_1\; \left({L\over L_0}\right)^{-\beta}\; , 
\label{eq:lumfunc}
\end{equation}
where $\alpha=1+\beta\neq 1$ and $N_1=N_0/(\alpha-1)$. In this
convention $L_0$ is a fiducial luminosity and $N_1$ and $N_0$ are
dimensionless constants (whose values depend on the choice of $L_0$).

The resulting index for spiral hosts, $\beta=0.8\pm 0.2$, is in
agreement with that found by Colbert et~al. (2004) for normal spiral
galaxies ($0.79\pm 0.24$). The cumulative luminosity function for the
sources in elliptical galaxies is best fitted by a broken power-law
with $\beta_1=0.40\pm 0.01$ (for $L < L_{\rm break}$), $\beta_2=1.4\pm
0.2$ (for $L > L_{\rm break}$) and $\log (L_{\rm break}/{\rm erg\;
s}^{-1})=38.7\pm 0.1$. The value of $\beta_2$ is comparable to the
value found by Colbert et al (2004) for normal elliptical galaxies
($1.41\pm 0.38$). But Colbert et~al. (2004) do not mention a break in
the luminosity function, although some of their luminosity functions
seem to flatten in their Figure 3. Other authors find a break in the
point-source luminosity function of normal elliptical galaxies
\citep{Sarazinal01, Bal01, Ral04}. On the other hand, Sivakoff et
al. (2003) do not find a break in the luminosity function of the two
elliptical galaxies NGC~4365 and NGC~4382. The break luminosity that
some authors find is comparable to that from our luminosity
function. This break could indicate that the nature of the XRB is
different for sources above and below the break luminosity. Sarazin
et~al. (2001) suggested that the accretor of XRBs below the break
luminosity is a neutron star, while the accretor of XRBs above the
break luminosity is a black hole, based on the agreement of the break
luminosity with the Eddington luminosity of a 1.4~\Msol\ neutron
star. The value of $\beta_1$ does not match any value cited by Colbert
et~al. (2004) even though it is closest to the slope of the cumulative
luminosity function of merger galaxies ($\beta=0.65$). At any rate,
the slope of the cumulative luminosity functions is well in the range
of those for normal galaxies so LINERs do not appear to have an
unusual XRB population.

NGC~3379 has 10 point sources in its inner 1 kpc region, so we were
able to construct a luminosity function for this individual galaxy. The
cumulative luminosity function is well fitted by a power-law with
$\beta=0.56\pm 0.03$. This value is close to the value found by
Colbert et al (2004) for starburst galaxies.

Using the above properties of the point source populations, we can
investigate whether the power-law component in the diffuse emission
spectra (see \S\ref{sec:diffuse}) can be due to the presence of
unresolved X-ray binaries. To test this hypothesis, we extrapolated
the luminosity function of equation~(\ref{eq:lumfunc}) to very low
luminosities and integrated it up to the minimum detectable luminosity
of a resolved point source, i.e.,
$$
L_{\rm XP} = \int^{L_2}_{L_1} L\; N(L)\; dL 
$$
$$
= {L_0\; N_1\; \beta\over 1-\beta}\; \left({L_2\over L_0}\right)^{1-\beta}
  \left[1-\left({L_1\over L_2}\right)^{1-\beta}\right]
$$
\begin{equation}
\approx {L_0\; N_1\; \beta\over 1-\beta} \left({L_2\over L_0}\right)^{1-\beta}\; ,
\label{eq:psource}
\end{equation}
where we have neglected the term in square brackets since $L_2 \gg
L_1$ and $\beta<1$.  We apply equation~(\ref{eq:psource}) to
elliptical and spiral galaxies using the lowest value from the
luminosity functions of Figure~\ref{fig:lumfunc} as the value of
$L_2$. We find that for elliptical galaxies ($L_0=10^{38}~{\rm
erg~s}^{-1}$, $N_1=1.3$, $L_2=6.6\times 10^{37}~{\rm erg~s}^{-1}$,
$\beta=0.82$), the estimated unresolved point source luminosity
(0.5--10~keV) is $L_{\rm XP} = 6\times 10^{38}~{\rm erg~s}^{-1}$,
while for spiral galaxies ($L_0=10^{38}~{\rm erg~s}^{-1}$, $N_1=54.3$,
$L_2=3.4\times 10^{37}~{\rm erg~s}^{-1}$, $\beta=0.40$), $L_{\rm XP} =
2\times 10^{39}~{\rm erg~s}^{-1}$. These luminosities are certainly
high enough to explain the hard X-ray component in the average diffuse
emission spectra Also the spectra of the hard component have a
power-law index of $\Gamma=1.2$, which is consistent with low hard
state X-ray binaries.


\section{Discussion of Individual Galaxies and Evaluation of the Power Source\label{sec:individual}}

From Figure~1 and Table 2, there is no obvious trend in the morphology
of the central kiloparsec of the galaxies. Some galaxies have a
central point source, some do not. Some galaxies have diffuse emission
and some do not. From Table 3 and Figure 2, the spectrum of the
brightest sources in the galaxies are not all fitted by the same model
and have a wide range of luminosity and model parameters. An
evaluation of the power source of each individual galaxy is then
required. To aid in our interpretation, we checked if the nuclei of
the galaxies were observed at other wavelengths, especially in the
radio and the UV. In general terms, a high X-ray radio loudness
parameter\footnote{This measure of radio loudness using X-ray
luminosity ($R_{\rm X}$) instead of the B luminosity ($R_{\rm O}$) was
introduced by Terashima \& Wilson (2003) and the cut-off value of
$-4.5$ was derived from the the optical cut-off $R_{\rm O} > 10$ and
the observed correlation between $R_{\rm X}$ and $R_{\rm O}$ for a
sample of LLAGNs, Seyferts and PG quasars: $\log R_O= 0.88\;\log
R_{\rm X} + 5.0$.}, defined as
\begin{equation}
\log R_{\rm X}= \log\left[
{\nu L_{\nu}{\rm (5~GHz)} \over L_{\rm X}{\rm (2-10\;keV)}}
\right] > -4.5
\label{eq:Rx}
\end{equation}
and a flat radio spectrum are indicative of a low luminosity AGN (hereafter LLAGN) \citep{TW03} so
the presence of a compact, high-brightness temperature radio core
provides strong evidence of the presence of a LLAGN. An unresolved UV
point source with non-stellar spectrum is also strong evidence for a LLAGN
\citep{Mal95}, while an extended UV source is unlikely to be an
AGN. If the flux of the UV point source varies over time, this
indicates an AGN \citep{Mal05}. When a broad H$\alpha$ was
detected (i.e., the LINER is classified as L1), this demonstratates
the presence of a LLAGN.

Finally, we checked whether our interpretation of the power source
could be supported by the estimate of the age of the stellar
population from population synthesis models, applied to specific
galaxies. In cases where an AGN is not a likely power source, the 
emission lines could be powered either by photoionization from 
massive stars from a young stellar population or from  post-AGB stars 
and the nuclei of young planetary nebulae associated with an 
old stellar population.

For LINERs with composite power sources (LLAGN plus stellar
processes), we determined the contribution of the AGN to the central
X-ray luminosity using the the spectrum from a circle of radius
2\farcs 5 at the center of the galaxy. We chose such an aperture
because it has a similar projected area as the aperture used in the
survey of Ho et al. (1997a) to measure the emission line
luminosities. A radius of 2\farcs 5 corresponds to 120~pc for a galaxy
at 10~Mpc and 300~pc for a galaxy at 25~Mpc, i.e., it represents the
very core the galaxies. We fitted this spectrum with a two-temperature
plasma model plus a power-law. The power-law component parameters were
fixed to those obtained during the analysis of the central point
source; the two-temperature plasma models represents the diffuse
emission. Then we calculated the 2--10 keV luminosity of
the central point source by integrating the power-law component, the
2--10 keV luminosity of the full 2\farcs 5 central region from the
two-temperature plasma plus power-law model, and thus the fraction of
the X-ray luminosity in the central region contributed by the point
source. The various 2--10~keV luminosities are listed in Table~4.
We use this fractional contribution as a basis for classifying a LINER
as a composite. When we use the 0.5--10 keV band to calculate the
contribution from the point source to the total central X-ray
luminosity, the numbers change slightly without changing the
conclusion about the dominant source (except in the case of NGC~4438
as explained in its individual description). All the luminosities are corrected for absorption.

\begin{description}

\item[NGC~1553] (L2/T2, S0) has three point sources in its central
kiloparsec. The brightest one is exactly at the center of the galaxy
and its spectrum is well fitted either by a two-temperature plasma
model ($kT_{1}=0.14^{+0.04}_{-0.09}$ keV, $kT_{2}=18.0^{+70}_{-10.4}$
keV) or a power-law ($\Gamma=1.2^{+0.2}_{-0.1}$). The preference of a
the two-temperature over a one-temperature plasma model and a power-law model is justified
by the F-test. Both of the above models give a luminosity
of $L_{\rm 0.5-10\; keV} = 1.5\times10^{40}$~{\rm erg~s}$^{-1}$. With
such a luminosity and spectral characteristics, this source is
probably an AGN. This is supported by observations in the radio band:
NGC~1553 is a weak radio source (10~mJy at 843~MHz; Harnett
1987). Assuming a radio spectral index of $p=-0.5$ (adopted as our
default value hereafter; where $S_{\nu}\propto \nu^p$), we calculate
the radio flux at 5 Ghz to be $S_{5\;{\rm Ghz}} = 4.1$ mJy and $\log
R_{\rm X} = -2.7$, which means that this source is radio
loud. Blanton,~Sarazin \& Irwin (2001) came to the same conclusion on
the nature of the central source using the same data set but they
estimated its luminosity to be slightly higher ($L_{\rm 0.3-10\;
keV}=1.75\times10^{40}$~{\rm erg~s}$^{-1}$).
%
%
We evaluated the luminosity of the central
2\farcs 5 region to be $L_{\rm 2-10\; keV} = 1.6\times 10^{40}$~{\rm
erg~s}$^{-1}$. The luminosity of the AGN is evaluated by
integrating the power-law that fits the data to be $L_{\rm 2-10\; keV}
= 1.5\times 10^{40}$~{\rm erg~s}$^{-1}$, which means that the AGN is
the source of 92\% of the X-ray luminosity from the central
region. This is consistent with estimates of the age of the stellar
population that indicate that the stars in the nucleus of NGC~1553 are
old ($>$ 17 Gyr, Terlevich \& Forbes 2002; 10-15 Gyr, Longhetti
et~al. 2000).

\item[NGC~2681] (L1, SB0/a) has three point sources within its central
kiloparsec. The brightest point source is at the exact center and is
detected in both the soft and hard bands. This is reminiscent of
NGC~1553, except that NGC~2681 has more diffuse emission than
NGC~1553. The spectrum of the central source is best fitted by a
two-temperature plasma model or a power-law plus a single-temperature
plasma model (see Table~3). 
%
%
The luminosity inferred from these models is $L_{0.5-10\;\rm keV} =
9.5\times10^{38}$~{\rm erg~s}$^{-1}$. These fits do not exclude the
possibility that the central source is a LLAGN since circumnuclear hot
gas can give rise to the thermal component superposed on a power-law
AGN spectrum. This source was detected in the radio waveband
($S_{\nu}$=12.0~mJy at 1.49~GHZ; Condon 1987) leading to $\log R_X =
-2.1$ (radio loud). A broad H$\alpha$ emission line was detected in
the optical \citep{Hal97b} suggesting that this source is a LLAGN. 
%
%
We estimated the total luminosity from the 2\farcs 5 central region to
be $L_{2-10\;\rm keV} = 9.7 \times 10^{38}$ {\rm erg s}$^{-1}$ and the luminosity of the AGN to be $L_{2-10\;\rm keV} = 1.8 \times
10^{38}$ {\rm erg s}$^{-1}$ (by integrating the power-law component),
which means that the potential AGN contributes only $\sim 20$\% of the
total X-ray luminosity of the central kiloparsec. This indicates that
the contribution from stellar processes is dominant. The other 80\% of
the luminosity comes from three other point sources in the central
2\farcs 5 region and diffuse emission. This is consistent with
estimates that the solar metallicity stellar population of NGC~2681
was formed in a starburst episode about 1 Gyr ago
\citep{GDal04,Cal01,Bal88}. An intermediate age ($10^8$--$10^9$ years)
population has also been detected in the nucleus of this galaxy by
\citep{CFal05}.

\item[NGC~3379] (T2, E1) harbors a cluster of 10 point sources in its
nucleus. One of these point sources (not the brightest one) is at the
exact center of the galaxy. The spectrum of this source had only 89
counts, thus its luminosity was estimated by fitting an
absorbed power-law: $L_{0.5-10\;\rm keV}=6.3\times 10^{37}$~{\rm
erg~s}$^{-1}$. An unresolved radio source of $S_{\nu}$=0.7~mJy has
been detected at 5~GHz at the center of this galaxy \citep{WH91}. We
calculate $\log R_X = -2.3$, which makes this source radio loud and
suggests that it is a LLAGN. 
%
%
The luminosity of the central 2\farcs 5 region was
evaluated to be $L_{2-10\;\rm keV} = 1.0\times10^{38}$~{\rm
erg~s}$^{-1}$ and the LLAGN has $L_{2-10\;\rm keV} =
1.7\times10^{37}$~{\rm erg~s}$^{-1}$. The candidate LLAGN contributes
$\sim$20\% to the luminosity of the central 2\farcs 5 region, making
this LINER a composite. However, age estimates indicate a stellar
population of intermediate age, in the range of 6--9 Gyr \citep{TF02,
Gal04}. We were able to fit the brightest source in the nucleus with a
number of models involving a power law and a soft component (diskbb,
bb, mekal). The addition of the soft component to the simple power-law
is not strongly justified by the F-test (chance probability between 14
and 44\% depending on the model for the soft component; a plasma model
to the power-law gives the smallest chance probability). The total
luminosity of this source is $L_{0.5-10\;\rm keV}=1.4\times
10^{39}$~{\rm erg~s}$^{-1}$. From its spectrum, luminosity, and
off-center position, one can infer that this source is probably an
X-ray binary (XRB).

\item[NGC~3507] (L2, SBb) has two point sources in its central
kiloparsec: one very faint (14 counts) close to the edge of the region
and a central extended source embedded in a bright patch of diffuse
emission. The central source had only 187 counts, which did not allow
us to test a wide variety of spectral models. Instead we estimated its
luminosity by fitting a simple mekal model (the photon index
of the best-fitting power-law model was unrealistic) and we find
$L_{0.5-10\;\rm keV} =1.2\times10^{39} ~{\rm erg~s}^{-1}$ after
correcting for absorption. A weak, {\it extended} radio source
($S_{\nu}$=16~mJy) is detected at 1.45~GHz \citep{C87}, which suggests
that this object is not a LLAGN. It could be a collection of SNRs
producing a superbubble around it seen as the X-ray diffuse emission
and radio extended source. The fact that the data could not be fitted
with a realistic power-law index strengthen this idea. This indicates
that the X-ray emission of this LINER is powered by stellar processes. This result is
consistent with the estimate of the age of the stellar population,
which indicates the presence of a young stellar component ($10^6$
years) \citep{GDal04}.

\item[NGC~3607] (L2, S0) has two point sources embedded in diffuse
emission, but none of them is at the center. The brightest point
source did not have enough counts to justify fitting models to its
spectrum; its luminosity was estimated to be $L_{0.5-10\;\rm keV}= 1.4
\times 10^{39}$~{\rm erg~s}$^{-1}$ by scaling the count rate to that
expected from a power-law model. The absence of a central point source
is a strong case against an AGN, in agreement with Terashima
et~al. (2002) who did not find evidence for an AGN from an X-ray
spectrum obtained with ASCA. The point sources and the diffuse
emission suggest that the X-ray emission is powered by stellar
processes. The age of the stellar population is estimated to be
between 3.6 and 5.6 Gyr \citep{TF02, PS02}.

\item[NGC~3608] (L2, E2) has two point sources with the brightest one
at the center, embedded in diffuse emission. The spectrum of the
brightest source does not have enough counts to fit a variety of
models to it but its luminosity was estimated to be $L_{0.5-10\;\rm
keV} = 8.7\times10^{38}$~{\rm erg~s}$^{-1}$. There is no detection of
this source at 5~GHz but only an upper limit of 0.5 mJy
\citep{WH91}. We can then put a limit on the radio loudness: $\log
R_{\rm X} < -2.7$. Thus this source might be radio loud, hence a
candidate LLAGN. We note that this galaxy is the second most distant
object in the sample and that the source is offset by 140\farcs  from
the aim point of the CCD and the 50\% encircled energy radius at
1.49~keV at this off-axis location is 1\farcs 5, i.e. a large fraction
of the 2\farcs 5 radius region used to determine the contribution to
the central X-ray luminosity. Also such a large PSF increases the
chance of source confusion and the chance that this is a multiple
source. The age of the stellar population in NGC~3608 has been
estimated to be between 9 and 10 Gyr \citep{TF02, PS02}.
We estimate that the luminosity of the central 2\farcs 5 is
$L_{2-10\;\rm keV} = 1\times10^{39}$~{\rm erg~s}$^{-1}$ and the
luminosity of the central source is $L_{2-10\;\rm keV} =
6\times10^{38}$~{\rm erg~s}$^{-1}$ so the hypothetical AGN contributes
60\% of the X-ray luminosity.

\item[NGC~3628] (T2, Sb) has five point sources in its central
kiloparsec including a bright point source 1~kpc away from the
center. There is no point source at the center of the galaxy, but a
central patch of diffuse emission. The spectrum of the diffuse
emission is well fitted by a variety of models with a two-temperature
plasma or a power-law giving the best fits. The resulting luminosity
is ($L_{0.5-10\;\rm keV}=4.6\times10^{38}$~{\rm erg~s}$^{-1}$). An
extended radio source was detected at the center of NGC~3628
\citep{C87}, which could indicate a SN-driven superbubble. This galaxy
is believed to be a starburst galaxy due to the presence of numerous
XRBs \citep{Sal01}. There is no study of the stellar population that
gives an estimate of the age. The absence of a LLAGN, the point
sources and the diffuse emission suggest that the X-ray emission of this LINER is powered by
stellar processes (perhaps a young starburst).

\item[NGC~4111] (L2, S0) has only one central point source with a hard
X-ray spectrum. There is also a fair amount of diffuse emission. The
spectrum extracted from the central source does not have enough counts
to justify fitting models to it but the luminosity can be
estimated to be $L_{0.5-10\;\rm keV} = 4.4\times10^{39}$~{\rm
erg~s}$^{-1}$ and $L_{2-10\;\rm keV} = 3.7\times10^{39}$~{\rm
erg~s}$^{-1}$. There is a detection of a central source in the radio with
$S_{5\;\rm GHz}=2.3$~mJy \citep{WH91}, which means that $\log R_{\rm
X} = -3$. The radio loudness, the central position of the source and
the hard X-ray spectrum suggest that this source is a LLAGN. 
%
%
We evaluated the luminosity of the central 2\farcs 5 region to be
$L_{2-10\;\rm keV} = 4.8\times10^{39}$~{\rm erg~s}$^{-1}$, meaning
that the point source contributes to $\sim 77$\% of the X-ray
luminosity. The age of the stellar population of NGC~4111 was
estimated to less than 1 Gyr \citep{G89}.

\item[NGC~4125] (T2, E6) has only one hard point source at its center
surrounded by diffuse emission. This point source did not have enough
counts to justify fitting various models to its spectrum, but its
luminosity was estimated to be $L_{0.5-10\;\rm keV}=9\times
10^{38}$~{\rm erg s}$^{-1}$. This source was not detected in the radio
but a limit was set: 0.5 mJy at 5 GHz (Wrobel\& Heeschen 1991). This
means that $\log R_{\rm X} < -2.7$ so this source could be radio
loud. Due to its central position and possible radio loudness, this
source could be a LLAGN. 
%
%
We estimate that the luminosity of the central 2\farcs 5
region is $L_{2-10~\rm keV}=6.7\times 10^{38}$~{\rm erg s}$^{-1}$ and
the LLAGN has $L_{2-10\;\rm keV}=5.4\times 10^{38}$~{\rm
erg s}$^{-1}$, i.e. $\sim 80$\% of the X-ray luminosity. The remaining
X-ray luminosity is contributed by the diffuse emission.

\item[NGC~4314] (L2, SBa) has no point sources, just sparse soft
diffuse emission. NGC~4314 is one of the nearest examples of galaxies
hosting a circumnuclear ring of star formation. Its diameter is
10\arcsec, which falls within the central kiloparsec. The age of the
stellar population in the nuclear ring has been estimated to be very
young (1--15 Myr, Benedict et~al. 2002). So the X-ray emission of this LINER is very likely
powered by young stars.

\item[NGC~4374] (L2, E1) has four point sources in its nucleus. The
brightest one is at the center and has a hard X-ray spectrum. There is
also significant soft, knotty, diffuse emission, although the knots
could be embedded point sources. There are many models that give good
fits, in particular a power-law and a plasma model plus a
power-law. The F-test justifies the addition of a plasma component to
the power-law (chance probability = 4$\times 10^{-4}$). The
luminosity is $L_{0.5-10\;\rm keV}=7.7\times 10^{39}$~{\rm erg
s}$^{-1}$. The stellar population of NGC~4374 is quite old (13.7 Gyr,
Proctor \& Sansom 2002; 11.8 Gyr, Terlevich \& Forbes 2002). Radio
images of NGC~4374 show structures resembling jets \citep{DYal80,
LB87}, which is a strong indication of an AGN. The flux of the core at
5 GHz was measured to $S_{\nu}=160$ mJy \citep{Nal02}, giving $\log
R_{\rm X}= -1.3$, which means that this source is radio-loud. 
%
%
The total X-ray luminosity in the central 2\farcs 5 radius region is
$L_{2-10~\rm{keV}}=4.4\times 10^{39}$~{\rm erg~s}$^{-1}$ and the
luminosity of the AGN is $L_{2-10~\rm{keV}}=4\times
10^{39}$~{\rm erg~s}$^{-1}$ (obtained by integrating the power-law
spectral component). So the contribution from the LLAGN to the X-ray
luminosity is 90\%.

\item[NGC~4438] (L1, S0/a) has only one point source, which is close
to the center but not exactly centered (within 2 $\sigma$). It has a
hard X-ray spectrum and is embedded in soft diffuse emission with an
elongated morphology. The spectrum of the central source is well
fitted by a thermal plasma model or a plasma component plus a
power-law with a total luminosity of $L_{0.5-10\;\rm keV}\sim7.7\times
10^{39}$~{\rm erg~s}$^{-1}$. Since the source is deeply embedded in
diffuse emission, it is possible that it is an AGN. This source is
weakly detected at 1.49~GHz with a flux of 116~mJy \citep{C87}, which
gives $\log R_{\rm X} < -1.2$, i.e. it can be radio loud. Also a broad
H$\alpha$ line was detected \citep{Hal97b}, which supports the AGN
hypothesis. An amorphous patch about 5\arcsec  in size was detected in
the UV \citep{Mal96}, which was interpreted as a superbubble and might
be associated with the extended patch of diffuse emission seen in the
X-ray. However, age estimates of the stellar population indicate that
an old population dominates the optical starlight spectrum in the
bulge \citep{CFal05,CFal04,BBal00}. 
%
%
We estimated the X-ray luminosity from the central 2\farcs 5 radius
region to be $L_{2-10\;\rm keV} = 1.4 \times 10^{39}$~{\rm
erg~s}$^{-1}$ and the luminosity of the AGN to be
$L_{2-10\;\rm keV} = 1.4 \times 10^{39}$~{\rm erg~s}$^{-1}$ (obtained
by integrating the power-law component). Thus the hard X-ray
luminosity of this LINER is primarily powered by the LLAGN. We note
that this is the only LINER whose dominant X-ray source luminosity
changes depending of what band we consider. The AGN dominates in the
2--10 keV band, while the plasma emission dominates in the 0.5--10 keV
band. 

\item[NGC~4457] (L2, S0/a) has one bright central source with a hard
spectrum, embedded in soft diffuse emission. The spectrum of the
central source is well fitted by a power-law plus plasma model or a
two-temperature plasma model ($L_{0.5-10\;\rm keV}=1.9\times
10^{39}$~{\rm erg~s}$^{-1}$). This can be interpreted as the spectrum
of an embedded AGN and is confirmed by the detection of this source at
1.49 GHz \citep{C87}. From this detection we infer $\log R_{\rm X} =
-1.6 $, which means that this source is radio loud.
%
%
The luminosity of the AGN is $L_{2-10\;\rm
keV}=9.7\times10^{38}$~{\rm erg s}$^{-1}$ while the luminosity of the
central 2\farcs 5 is $L_{2-10\;\rm keV} = 1.8\times 10^{39}$~{\rm erg
s}$^{-1}$ so the AGN contributes $\sim$ 55\% of the X-ray
luminosity. This rest of the luminosity comes from the diffuse
emission, so this LINER is a composite.

\item[NGC~4494] (L2, E1-2) has six point sources in its central
kiloparsec and the coordinates of the brightest source agree with the
center of the galaxy within 2 $\sigma$. This source has a hard X-ray
spectrum, which is best fitted by a two-temperature plasma model or a
power-law. The luminosity is $L_{0.5-10\;\rm keV}=1.6\times
10^{39}$~{\rm erg s}$^{-1}$. There is no detection of a source at
5~GHz, with an upper limit of 0.5 mJy \citep{WH91}, leading to $\log
R_{\rm X} < -3.7$. Thus, this source could be radio loud and could be
an AGN. 
%
%
The luminosity of the central 2\farcs 5 is $L_{2-10\;\rm keV} =
1.1\times 10^{39}$~{\rm erg s}$^{-1}$ and the luminosity of the power AGN is $L_{2-10\;\rm keV} = 9.9\times 10^{38}$~{\rm
erg s}$^{-1}$ so the AGN candidate contributes 90\% of the
luminosity. The stellar population was estimated to be less than 5 Gyr
old \citep{OSP04}.

\item[NGC~4552] (T2, E1) has a bright, hard central source, plus four
other point sources in its central kiloparsec and soft diffuse
emission. The spectrum of the central source is well fitted by a
power-law or a comptonized blackbody. These models give a luminosity
of $L_{0.5-10\;\rm keV}=3.7\times 10^{39}$~{\rm erg s}$^{-1}$. This
source was clearly detected at 4.85~GHz with $S_{\nu}=64~$mJy and
a spectral index $p=-0.23$ \citep{Cal91} so we calculate $\log R_{\rm
X} = -1.5$, which means that this source is radio loud. Moreover a UV
flare from this source was detected \citep{Ral95,Cal99} and the UV
flux was observed to vary over a short period of time \citep{Mal05}
indicating a weak AGN.  The stellar population was estimated to be old
(9.6 Gyr, Terlevich \& Forbes 2002). 
%
%
The X-ray luminosity from the central 2\farcs 5 region is
$L_{2-10\;\rm keV}=3.4\times 10^{39}$~{\rm erg s}$^{-1}$ and the
luminosity of the AGN is $L_{2-10~\rm keV}=2.0\times
10^{39}$~{\rm erg s}$^{-1}$. Thus the X-ray luminosity of this LINER
comes mostly from its AGN ($\sim 60\%$).

\item[NGC~4636] (L1, E1/S0) has bright knotty, soft, diffuse
emission. There are five point sources deeply embedded in the diffuse
emission. One of them is at the center but this is not the brightest
source and its spectrum is soft. This source did not have enough
counts to justify fitting its spectrum with a variety of models, but
its luminosity was evaluated to be $L_{0.5-10\;\rm keV}=8.7\times
10^{38}$~{\rm erg s}$^{-1}$. A broad H$\alpha$ component was detected
\citep{Hal97b} indicating a LLAGN, but the X-ray signature of an AGN
is not detected. Loewenstein et~al. (2001) failed to detect an AGN
using the same data set, as did Terashima et~al. (2002) using ASCA
observations. If there is an AGN (as suggested by the detection of a
broad H$\alpha$line ), it could be heavily absorbed.

\item[NGC~5055] (T2, Sb/c) has a five point sources in its nucleus,
clustered close to the center, and a bright hard source at the very
center. The spectrum of this source is well fitted by a
two-temperature plasma model, but a power-law plus plasma model
gives a very poor fit. We did not attempt more complex models due to
the small number of degrees of freedom. Maoz et~al. (1995) resolved
the central point source in the UV and obtained a radius of 7~pc (at
the distance assumed in this paper). Maoz et~al. (2005) did not
observe any variation of the UV flux of this source. All these pieces
of evidence indicate that this source is more likely a compact star
cluster. This source is not detected at various radio frequencies
except at 57.5~MHz with $S_{\nu}$=2.1~Jy \citep{IM90}, which
means that this source is radio quiet ($R_X=-6$) and not likely to be
an AGN. The stellar population of NGC~5055 has a range of ages from 1
Myr to 10 Gyr with most of the population of intermediate and old age
\citep{GDal04, CFal04}. So the X-ray emission of this LINER is
powered by stellar processes.

\item[NGC~5866] (T2, S0) has a central, hard X-ray source and one other
point source far away from the center and embedded in soft diffuse
emission. The spectrum of the central source did not have enough
counts to attempt a fit. However its luminosity was estimated to be
$L_{0.5-10\;\rm keV}=3.2\times 10^{38}$ erg s$^{-1}$ with 98\% of it
emitted in the hard band. This is also the luminosity from the central
2.5'' region indicating that this source is the main contributor to
the central X-ray luminosity. This source has $S_{\nu}=7.5$ mJy at 5
GHz \citep{Nal00}. We calculated that $\log R_{\rm X}=-1.5$, which means that
this source is radio loud and possibly a LLAGN.

\item[NGC~7331] (T2, Sb) has three point sources but none are at the
center (at the $3\sigma$ confidence level). There is no hard X-ray
emission close to the center of the galaxy. No source had enough
counts to justify spectral fits. The stellar population does not have
a young component and is mostly 1-10 Gyr old
\citep{CFal05,CFal04,GDal04}. So this LINER might be powered by
stellar processes, most likely associated with the old stellar
population.

\end{description}

\section{Summary of Results and Discussion\label{sec:discussion}}

\subsection{Summary and Immediate Implications for AGN Demographics}

In order to generalize our results, the first issue we examine is
whether the present sample is representative of the entire population of
LINERs, in terms of the distribution of LINER classes. The ratio of
transition objects to pure LINERs in the Ho et~al. (1997c) survey is
0.75 while in our sample, it is 0.63 (ignoring NGC~1553, which was not
classified by Ho). So the transition objects are slightly
underrepresented in our sample. Moreover in our sample, 27\% of the
pure LINERs are L1, which is comparable to the 25\% of L1 LINERs in
the Ho et al. (1997c) survey.  We also note that the H$\alpha$
luminosity distribution of the sample closely resembles that from
the Ho et al (1997c) sample. Thus, one can generalize the results from
this sample with reasonable confidence.

Soft, circumnuclear diffuse emission is common in our objects with a
typical 0.5--2~keV luminosity on the order of $10^{38}~{\rm
erg~s}^{-2}$, in agreement with the results of Gonz\'alez-Martin et
al. (2005, 2006). In 12 out of 19 LINERs there are signs of an AGN
whether it dominated the X-ray luminosity or not. In 5/19 LINERs we
did not detect an AGN, with upper limits to the luminosity as listed
in Table~4.  For 2/19 LINERs (NGC~4636 and NGC~7331), it was difficult
to determine whether there was an AGN because of the knotty structure
of the diffuse emission. Thus, at least 63\% (and possibly up to 74\%)
of the objects in the sample harbor an AGN.

Table 4 summarizes the 0.5--2 and 2--10 keV luminosity of the AGN
candidates in each galaxy (Columns 6 and 7) along with their Eddington
ratio (Column 8; see details below) and X-ray radio loudness parameter
(Column 9). When comparing the 2--10 keV luminosity of the AGN
(Column 4) to the luminosity of the diffuse emission (Columns 2 and
3), we notice that for 10 of the 12 galaxies with an AGN candidate,
the AGN dominates the 2--10~keV luminosity in the inner {2\farcs}5; in
9 out of these 10 cases, the AGN dominates the total 2--10~keV
luminosity in the inner kiloparsec. When we consider the luminosity in
the entire 0.5--10~keV band, however, the contribution of the diffuse
emission is often comparable to the contribution of the AGN, even in
the inner {2\farcs}5. Figure 5 shows the distribution of luminosities
of the AGNs found in our sample (and upper limits when there is no
detection of an AGN), the distribution of $\log R_{\rm X}$ and the
distribution of the fractional contribution of the AGN to the 0.5--10
keV luminosity of the central 2\farcs 5 region. We note that the
detected AGNs span a wide range in 2--10 keV luminosity, from
$1.7\times 10^{37}$ to $1.2\times 10^{40}~{\rm erg~s}^{-1}$, with an
average of $L_{\rm 2-10\; keV}=2.3\times10^{38}~{\rm
erg~s}^{-1}$. This luminosity range is comparable to that found by
Terashima \& Wilson (2003) for LLAGN in LINERs with bright, compact
radio sources.

We can express the Eddington ratio of the AGNs we have detected (the
ratio of the bolometric luminosity of the AGN and the Eddington
luminosity) as follows:
$$
\REdd\equiv{L_{\rm bol}\over L_{\rm Edd}} = 
{L_{\rm 2-10\; keV}/\zeta 
\over 1.3\times 10^{38}\; (M_{\rm BH}/\Msol)~{\rm erg~s}^{-1}}
$$
\medskip
\begin{equation} 
= 7\times 10^{-7}\; L_{38}\; \zeta_{-1}^{-1}\; M_7^{-1}\; , 
\label{eq:eddington}
\end{equation}
where $M_{BH}=10^7 M_7~\Msol$ is the black hole mass reported in
Table~1, $L_{\rm 2-10\; keV} = 10^{38}\; L_{38}~{\rm erg~s}^{-1}$ is
the X-ray luminosity (reported in Table~4) and $\zeta =
0.1\,\zeta_{-1}$ is a ``bolometric correction'' (see Ho 1999).  The
resulting values of $\REdd$ are included in column 8 of Table~4;
they span the range $-8.2 < \log \REdd < -5.1$, for {\it detected}
AGNs in our sample.  These values are comparable to those found by
Filho et~al. (2004), for a sample of transition-type LINERs (using the
same method for black hole mass determination). These results however
are significantly different from those of Satyapal et al. (2005) who
find $-6.7 < \log\REdd < -0.3$ and $2\times 10^{38} < L_{2-10\rm{
keV}} < 10^{42} \rm{ erg s}^{-1}$ for the candidate AGNs in their
infrared-bright LINER sample. We note that they used a bolometric
correction three times higher than the one we use, but this does not
account for the whole difference. Their AGNs have higher X-ray
luminosities so the very low Eddington ratios AGNs in our sample are
probably the very low luminosity AGNs missing from their sample.

\subsection{Comparison With Normal Galaxies and the Source of Fuel for the AGN}

The X-ray luminosities and Eddington ratios of the accreting black
holes in our sample of LINERs are very similar to those of black holes
in the ``quiescent'' galaxies studied by Soria et
al. (2006a,b). Moreover, the properties of the diffuse emission (gas
temperature, density, and soft X-ray luminosity; see
\S\ref{sec:diffuse}) as well as the black hole masses in the two
samples are quite similar. Therefore, we can infer that the Bondi
accretion rates should also be similar.  Indeed, an inspection of
Figure~5 of Soria et al. (2006b) shows that several of the LINERs in
our sample fall in the same region of the Eddington ratio $vs$ Bondi
rate diagram. As Soria et al. (2006b) demonstrate, the fuel necessary
to power the observed low-level activity in their sample can easily be
provided by a combination of mass loss from stars within the black
hole's ``sphere of influence'' and from the flow of hot interstellar
gas into that sphere from larger radii (Bondi accretion). In fact, the
hot interstellar gas itself is likely to have originated in stellar
mass loss over a larger volume around the center of the galaxy (see
our earlier discussion in \S\ref{sec:diffuse}). The same
interpretation applies to the LINERs that we study here, based on the
close similarities between the nuclear properties of the two samples.

These similarities also contribute to the puzzle of why some galaxies
host LINERs while others do not. To address this question we examine
the energy budget of the AGN in the next section in the broader
context of the power-source behind the optical emission-line spectra of
LINERs.

\subsection{The Role of the AGN in the Grand Scheme of Things}

In this sample, stellar processes, as manifested by soft, diffuse
emission, seem to frequently contribute to the energy budget of the
central 2\farcs 5 region, whether alone or in combination with an
AGN. The AGN is often not the major power source. To illustrate this
we plot in Figure~\ref{fig:balance} the luminosity from the diffuse
emission $vs$ the luminosity from the AGN in the 0.5--10~keV and
2--10~keV bands. In more than half of the cases the diffuse emission
dominates.  It would be interesting to determine whether the dominant
contribution to the X-ray luminosity depends on the LINER
classification or the Hubble type, but our sample is too small to
allow any firm conclusions on these questions.

To investigate the role of the AGN in powering the observed optical
emission lines, we looked for correlations between the different X-ray
luminosities and the H$\alpha$ luminosity (taken from Ho et al 1997a).  We
find a correlation between $L_{\rm H\alpha}$ and $L_{\rm 2-10 keV}$ of
the central {2\farcs}5 region (Spearman rank correlation coefficient
$r_{\rm s}=0.33$). Our objects fall in the same region of the
$L_{\rm\; 2-10 keV}$--$L_{\rm H\alpha}$ diagram as the
lowest-luminosity objects of Ho et al. (2001) and also very close to
the correlation between these two quantities followed by low-redshift
quasars and Seyfert 1 galaxies. This can be seen in
Figure~\ref{fig:hoandmine} where the squares represent the data from
our sample and the stars represent the LINERs observed by Ho
et~al. (2001)\footnote{Ho et al. (2001) scaled a power-law model with
$\Gamma=1.8$ and $N_{\rm H}=2\times 10^{20}$ cm$^{-2}$ to the observed
count rate of the central sources to estimate the luminosity of the
sources. The objects from Ho et al. (2001) that do not follow the
correlation are mostly transition-type LINERs, which were not included
in the calculation of the correlation.}. We also draw the best fit
found by Ho et~al. (2001) as the solid line and the range of
luminosity where a starburst would fall (dotted lines).  The latter
locus is obtained by combining relations between these two quantities
and the star formation rate (Kennicutt 1998; Ranalli, Comastri, \&
Cetti 2003; Grimm, Gilfanov, \& Sunyaev 2003; Colbert et~al. 2004;
Persic et~al. 2004). It is noteworthy that none of our objects are
close to the starburst locus, which is consistent with an absence of
young stellar populations in the majority of the galaxies in our
sample.

In Figure~\ref{fig:lxvslha}, we focus on the correlation between the
{2\farcs}5 X-ray and H$\alpha$ luminosities for the objects in our
sample.  The trend persists regardless of whether we use the 2--10~keV
or the 0.5--2~keV luminosity. In fact, the correlation appears
stronger when considering the soft X-ray band instead of the hard
X-ray band: the Spearman rank correlation coefficients are 0.48 and
0.33 in the soft and hard bands respectively. This suggests that
production mechanism of the H$\alpha$ line is not strongly correlated
to the source of the hard X-ray radiation.

Figure~\ref{fig:lxvslha} gives a preliminary indication that
photoionization by the AGN is unlikely to be responsible for powering
the optical emission lines. We have carried out the following tests to
investigate this issue further.

\begin{enumerate}

\item
The first test is based on energy balance between the emission lines
and the ionizing continuum. Assuming that the ionizing continuum is a
power-law of photon index $\Gamma$ from 1~Ry to 100~keV (consistent
with ADAF models), its total luminosity can be obtained from the
observed 2--10~keV luminosity via $L_i = I(\Gamma)\; L_{\rm 2-10\;
keV}$, where the function $I(\Gamma)$ has values between 5 and 12 for
$\Gamma$ in the range 1.1--1.9. Thus for the ionizing continuum to
power the H$\alpha$ luminosity alone, we require that $L_{\rm 2-10\;
keV} > 5\; L_{\rm H\alpha}$. Out of the 19 objects in our sample, only
four satisfy this condition, under the most optimistic assumption. In
the case of NGC~1553, NGC~4374 NGC~4494, and NGC~4552 the ionizing
continuum luminosity exceeds the H$\alpha$ luminosity by a factor of a
few in the best case. Additional power would be available, if the AGN
SED included a significant ``UV bump'', but this appears rather
unlikely, in view of the LLAGN SED shapes shown by Ho (1999). It is
also unlikely that we have underestimated the X-ray luminosity of the
AGN because of absorption. This is because the spectra of 2/3 of our
AGN have a sufficiently high S/N that we were able to fit them,
determine the absorbing column, and correct the luminosity (of
course, we cannot rule out the possibility that the entire X-ray
emission is scattered and that we do not have a direct view of the
AGN).

\item
Another test for the photoionization hypothesis relies on photon
counting. Assuming the same continuum shape as above, we can express
the ionizing photon rate produced by the AGN in terms of the observed
X-ray luminosity as $Q_i = 6.25\times 10^{47}\; G(\Gamma)\; (L_{\rm
2-10\; keV}/10^{39}\;\ergs)~{\rm s}^{-1}$. The function $G(\Gamma)$
has values between 2 and 30 for $\Gamma$ in the range 1.1--1.9.  By
relating this to the H$\alpha$ photon rate, we can derive the
following requirement for the H$\alpha$ luminosity: $L_{\rm H\alpha} <
0.007\; f\; G(\Gamma)\; L_{\rm 2-10\; keV}$, where $f$ is the fraction
of ionizing photons absorbed by the emission-line gas and one in 2.2
recombinations leads to the emission of an H$\alpha$ photon
(Osterbrock 1989; we have assumed that there are no secondary
ionizations or excitations by fast photoelectrons). None of the
objects in our sample satisfy this condition, even if $f=1$. The four
objects identified in the previous paragraph could satisfy this
condition if each ionizing photon leads to 2--5 ionizations, on
average. All other objects would require more than 10 secondary
ionizations; most require more than 100.

\end{enumerate}

Thus the AGNs in the objects of our present sample are not
energetically significant. It is also noteworthy that in more than
half the cases, the electromagnetic power of the AGN is not enough to
drive the diffuse X-ray emission. Examining the samples of Terashima
\& Wilson (2003) and Filho et al. (2004), we notice a similar
effect. In particular, the former sample includes AGNs typically an
order of magnitude more luminous than ours but the faintest 25\% of
their AGNs fail the energy budget test and cannot power the H$\alpha$
emission. In the latter sample the X-ray detected AGNs have comparable
luminosities to ours and they all fail the energy budget
test.\footnote{~Filho et al. (2004) determine the X-ray luminosities
without fitting for the obscuring column density but they attribute
the low X-ray luminosity to absorption. Their three longest
observations are analyzed here as well.  In NGC~4552 we find that
intrinsic luminosity of the AGN, after correcting for absorption, is
not enough to balance the energy budget. In NGC~5866 there are too few
counts to warrant a fit, and in NGC~7331, we do not detect an AGN at
all.}  It appears, at least at first glance, that as the luminosity of
the AGN decreases, its important as the power source of the optical
emission lines diminishes too.  Therefore, we conclude our discussion
by considering other possible sources of power for the optical
emission lines below.

\subsection{Epilogue: What Is the Power Source of the Optical Emission Lines?}

The analysis presented above shows that the properties of the LINERs
in the present sample are very similar to those of normal
galaxies. More specifically, the hot interstellar gas in these LINERs
has, on average, the same temperature, density, and luminosity as in
normal galaxies. Moreover, the average properties of the stellar
populations in these LINERs are similar to those of normal galaxies:
the X-ray binary luminosity functions are the same and in most of our
LINERs the optical spectra indicate old stellar populations.  Finally,
even though an AGN is often detected, its electromagnetic power does
not appear to be significant.  Thus we must look for an explanation
for the LINER optical spectra that involves either stellar processes
or a mechanical interaction of the AGN with its immediate environs.

In three of the LINERs in our sample (NGC~3507, NGC~3628, and
NGC~5055) there is evidence for a young stellar population, while in
one other object (NGC~2681) there is evidence for an intermediate-age
population. In more general terms, Cid Fernandes et al. (2005) find
that 38\% of LINERs and transition objects have intermediate-age
stellar populations in their very nuclei. Hot stars from a young
stellar population are obvious sources of ionizing photons that could
power the optical emission lines (see \S1). In old and especially in
intermediate-age stellar populations, ionizing photons can be provided
by post-AGB stars and the central stars of planetary nebulae
\citep{Bal94}.

An alternative explanation is that the AGN powers the emission lines
by depositing mechanical energy via outflows (jets or winds) into the
emission line gas. At the low Eddington ratios that we find here, the
kinetic power of jets from the AGN can be significantly higher (by up
to 3 orders of magnitude) than the electromagnetic luminosity (see
Nagar, Falcke, \& Wilson 2005). Some of this power may be dissipated
by shocks in the immediate vicinity of the AGN, which would naturally
lead to LINER-like emission line spectra. This hypothesis is bolstered
by the observation that the emission-line regions of LINERs are rather
amorphous or filamentary on scales of 10--100~pc \citep{Pogge00}. This
type of feedback from the AGN may also contribute to additional
heating of the diffuse gas, which in turn would limit the accretion
rate and lead to the very low Eddington ratio that we infer (see the
discussion by Soria et al. 2006b).

In conclusion, we note that even though some LINERs are obviously
powered by their AGNs (M81 and NGC~4579 are among the best known
examples), the electromagnetic luminosity of the AGN is not always
enough to account for the observed emission-line luminosity. All of
the physical processes discussed in the literature so far may
contribute to powering the emission line spectra, making LINERs a
heterogeneous class of objects and suggesting that AGN photoionization
may be the exception rather than the rule.

\acknowledgements

This work was supported by the National Aeronautics and Space
Administration through {\it Chandra} award number AR4-5010A issued by
the {\it Chandra} X-Ray Observatory Center, which is operated by the
Smithsonian Astrophysical Observatory for and on behalf of the
National Aeronautics and Space Administration under contract
NAS8-03060. We would like to thank Patrick Broos and Leisa Townsley
for their assistance with the use of the {\tt ACIS Extract} software.
We also thank the anonymous referee for helpful comments.

\singlespace

\begin{deluxetable}{cccrccccc}
\tabletypesize{\small}
\tablenum{1}
\tablewidth{0in}
\tablecolumns{7}
\tablecaption{Target Properties and Observation Details}
\tablehead{
\colhead{Galaxy\tablenotemark{a}}      &
\colhead{LINER}                        &
\colhead{Hubble}                       &
\colhead{Dist.\tablenotemark{b}}       & 
\colhead{$N_{\rm H}$\tablenotemark{c}} & 
\colhead{$L_{\rm H\alpha}$\tablenotemark{d}} &
\colhead{}                             &
\colhead{Exposure}                     &
\colhead{Observation}                  \\
\colhead{(NGC)}                        &
\colhead{Type\tablenotemark{b}}        &
\colhead{Type\tablenotemark{b}}        &
\colhead{(Mpc)}                        &
\colhead{($10^{20}~{\rm cm}^{-2}$)}    &
\colhead{(erg s$^{-1}$)}               &
\colhead{$\log\left({M_{\rm BH}\over\Msol}\right)$\tablenotemark{e}} &
\colhead{(ks)}                         &
\colhead{Date}                         
}
\startdata
1553 &L2/T2\tablenotemark{f}& S0    & 16.1 & 1.50 & $6.61\times 10^{38}$ & 8.0   & 34.17 & 2000/01/02 \\ 
2681 &  L1 & SB0/a & 13.3 & 2.42 & $6.76\times 10^{38}$ & 7.2   & 80.00 & 2001/01/30 \\ 
3379 &  T2 & E1    &  8.1 & 2.75 & $8.71\times 10^{37}$ & 8.3   & 31.92 & 2001/02/13 \\ 
3507 &  L2 & SBb   & 19.8 & 1.63 & $2.45\times 10^{39}$ & \dots & 39.76 & 2002/03/08 \\ 
3607 &  L2 & S0    & 19.9 & 1.48 & $8.51\times 10^{38}$ & 8.4   & 39.00 & 2001/06/12 \\ 
3608 &  L2 & E2    & 23.4 & 1.49 & $1.91\times 10^{38}$ & 8.1   & 39.00 & 2001/06/12 \\ 
3628 &  T2 & Sb    &  7.7 & 2.23 & $7.41\times 10^{36}$ & 7.9   & 58.70 & 2000/12/02 \\ 
4111 &  L2 & S0    & 17.0 & 1.40 & $2.51\times 10^{39}$ & 7.8   & 15.00 & 2001/04/03 \\ 
4125 &  T2 & E6    & 24.2 & 1.84 & $9.12\times 10^{38}$ & 8.4   & 65.08 & 2001/09/09 \\ 
4314 &  L2 & SBa   &  9.7 & 1.78 & $2.82\times 10^{38}$ & 7.2   & 16.28 & 2001/04/02 \\ 
4374 &  L2 & E1    & 16.8 & 2.60 & $7.76\times 10^{38}$ & 8.8   & 28.85 & 2000/05/19 \\ 
4438 &  L1 & S0/a  & 16.8 & 2.66 & $2.34\times 10^{39}$ & \dots & 25.40 & 2002/01/29 \\ 
4457 &  L2 & S0/a  & 17.4 & 1.80 & $3.72\times 10^{39}$ & 7.0   & 39.38 & 2002/12/03 \\
4494 &  L2 & E1-2  &  9.7 & 1.52 & $3.47\times 10^{37}$ & 7.6   & 25.16 & 2001/08/05 \\ 
4552 &  T2 & E1    & 16.8 & 2.57 & $3.31\times 10^{38}$ & 8.5   & 55.14 & 2001/04/22 \\ 
4636 &  L1 & E1/S0 & 17.0 & 1.81 & $1.86\times 10^{38}$ & 8.2   & 53.05 & 2000/01/26 \\ 
5055 &  T2 & Sb/c  &  7.2 & 1.32 & $8.13\times 10^{37}$ & 7.1   & 28.36 & 2001/08/27 \\ 
5866 &  T2 & S0    & 15.3 & 1.47 & $1.02\times 10^{38}$ & 7.7   & 34.18 & 2002/11/14 \\
7331 &  T2 & Sb    & 14.3 & 8.61 & $3.09\times 10^{38}$ & 7.6   & 30.13 & 2001/01/27 \\
\enddata
\tablenotetext{a\;}{The coordinates of the nucleus are from isophotal 
contour fits to 2MASS images, with the following exceptions: 
NGC~4374 from VLBI observations \citep{Fal04},
NGC~4438 from VLA observations  \citep{HS91},
NGC~4552 from VLBA observations \citep{Nal02},
NGC~7331 from stellar kinematics measurements \citep{AC90}.}
\tablenotetext{b\;}{LINER types, galaxy morphological types, and
distances taken from Ho et~al. (1997c). LINER types are explained in
\S\ref{sec:targets} of the text.}
\tablenotetext{c\;}{Galactic \ion{H}{1} column densities from Dickey
\& Lockman 1990}
\tablenotetext{d\;}{H$\alpha$ luminosity, taken from Ho et
al. (1997a), except NGC~1553, which was taken from Phillips et
al. (1986)}
\tablenotetext{e\;}{Black hole mass based on the stellar velocity
dispersion.  See \S\ref{sec:targets} of the text.}
\tablenotetext{f\;}{The emission line strengths reported by
Phillips et al. (1986) and Veron-Cetty \& Veron (1986) indicate 
an L or T classification. In the absence of information on 
broad Balmer lines, we tentatively list this object as L2/T2.}
\end{deluxetable}

\begin{deluxetable}{ccccc}
\tablenum{2}
\tablewidth{0in}
\tablecolumns{6}
\tablecaption{Morphology of the Central Kiloparsec of the Target Galaxies}
\tablehead{
\colhead{}             &
\colhead{Number}       &
\colhead{Central}      &
\colhead{Diffuse}      &
\colhead{Point Source} \\
\colhead{Galaxy}       &
\colhead{of Point}     &
\colhead{Point Source} &
\colhead{Emission}     &
\colhead{Sensitivity\tablenotemark{b}}   \\
\colhead{(NGC)}        &
\colhead{Sources}      &
\colhead{Properties\tablenotemark{a}}&
\colhead{Properties\tablenotemark{a}}&
\colhead{(erg s$^{-1}$)} 
}
\startdata
1553  &   3  &  HS  &     &  $2.6\times10^{37}$ \\
2681  &   3  &  HS  &  S  &  $7.6\times10^{36}$ \\
3379  &  12  &  S   &     &  $7.0\times10^{36}$ \\
3507  &   2  &  S   &  S  &  $3.4\times10^{37}$ \\
3607  &   2  &      &     &  $3.4\times10^{37}$ \\
3608  &   2  &  S   &  S  &  $4.8\times10^{37}$ \\
3628  &   5  &      &  H  &  $3.4\times10^{36}$ \\
4111  &   1  &  H   &  S  &  $6.6\times10^{37}$ \\
4125  &   1  &  HS  &     &  $3.0\times10^{37}$ \\
4314  &   0  &      &  S  &  $1.9\times10^{37}$ \\
4374  &   4  &  HS  &  S  &  $3.3\times10^{37}$ \\
4438  &   1  &  HS  &  S  &  $3.8\times10^{37}$ \\
4457  &   1  &  HS  &  S  &  $2.6\times10^{37}$ \\
4494  &   6  &  S   &     &  $1.3\times10^{37}$ \\
4552  &   5  &  HS  &  S  &  $1.7\times10^{37}$ \\
4636  &   5  &  S   &  H  &  $1.8\times10^{37}$ \\
5055  &   5  &  HS  &     &  $6.2\times10^{36}$ \\
5866  &   2  &  H   &  S  &  $2.3\times10^{37}$ \\
7331  &   3  &      &     &  $2.3\times10^{37}$ \\
\enddata

\tablenotetext{a\;}{H indicates that the source is detected in the
hard (2--10 keV) band and S indicates the source is detected in the
soft (0.5--2 keV) band.}

\tablenotetext{b\;}{The lowest detectable 0.5--10~keV point source
luminosity in a given observation (assuming an absorbed power-law
spectrum with $\Gamma = 1.8$, and $N_{\rm H}=2\times 10^{20}~{\rm
cm}^{-2}$; see \S\ref{sec:fits}).}

\end{deluxetable}

\begin{deluxetable}{llcc}
\rotate
\tablenum{3}
\tablewidth{0in}
\tablecolumns{4}
\tablecaption{Spectral Models and Parameters for Bright Sources}
\tablehead{
\colhead{}                                  &
\colhead{}                                  &
\colhead{}                                  &
\colhead{Unabsorbed}                        \\
\colhead{Galaxy}                            &
\colhead{}                                  &
\colhead{}                                  &
\colhead{$L_{\rm 0.5-10\; keV}$}            \\
\colhead{(NGC)}                             &
\colhead{Best-Fitting Models and Parameters\tablenotemark{\;\rm a}}&
\colhead{$\chi^2$/d.o.f.}                   &
\colhead{(erg s$^{-1}$)}
}
\startdata
\noalign{\vskip -12pt}
\sidehead{\underline{\it Central Point Sources}}
1553 & mekal2T,~~ $N_{\rm H}=5^{+2}_{-1}\times 10^{21}$ cm$^{-2}$, $kT_1=0.14^{+0.04}_{-0.09}$ keV, $kT_2=18^{+70}_{-10}$ keV & 51.76/40 & $1.5\times10^{40}$\\
     & pow,~~ $N_{\rm H}=3.2^{+0.8}_{-0.7}\times 10^{21}$ cm$^{-2}$, $\Gamma=1.2^{+0.2}_{-0.1}$ & 63.6/43 & $1.6\times10^{40}$ \\
2681 & mekal2T,~~ $N_{\rm H}=2.3^{+2.0}_{-0.8} \times 10^{21}$ cm$^{-2}$, $kT_1=0.64^{+0.11}_{-0.07}$ keV, $kT_2=4^{+96}_{-2}$ keV & 8.3/14 & $9.0\times10^{38}$\\
     & pow+mekal,~~ $N_{\rm H}=2.7^{+0.6}_{-0.9}\times10^{21}$ cm$^{-2}$, $kT=0.64^{+0.10}_{-0.08}$ keV, $\Gamma$=2.0$^{+0.8}_{-0.9}$ & 8.2/14 & $9.9\times10^{38}$\\
4374 & mekal+pow,~~ $N_{\rm H}=2.0^{+7.9}_{-0.6}\times 10^{21}$ cm$^{-2}$, $kT=0.6^{+0.5}_{-0.4}$ keV, $\Gamma=2.0^{+0.2}_{-0.1}$ & 14.6/27 & $7.7\times 10^{39}$\\
     & pow,~~ $N_{\rm H}=1.9\pm 0.5\times 10^{21}$ cm$^{-2}$, $\Gamma=2.1\pm 0.2$ & 18.4/29 & $7.5\times10^{39}$\\
4438 & mekal+pow,~~ $N_{\rm H}=1.2\pm 0.9\times 10^{21}$ cm$^{-2}$, $kT=0.8\pm0.6$ keV, $\Gamma=1.1\pm1.1$ & 9.3/12 & $4.0\times10^{39}$\\ 
     & mekal,~~ $N_{\rm H}=3^{+2}_{-3}\times 10^{21}$ cm$^{-2}$, $kT=0.8^{+0.1}_{-0.2}$ keV & 8.4/14 & $9.4\times10^{38}$\\ 
4457 & mekal+pow,~~ $N_{\rm H}=9.8\pm 0.1\times 10^{20}$ cm$^{-2}$, $kT=0.6\pm 0.8$ keV, $\Gamma=1.7\pm 0.3$ & 6.1/9 & 2.1$\times 10^{39}$\\
     & mekal2T,~~ $N_{\rm H}=8.^{+8}_{-6}\times 10^{20}$ cm$^{-2}$, $kT_1=0.7^{+0.3}_{-0.2}$ keV, $kT_2=7^{+42}_{-3}$ keV & 8.8/9 & 1.8$\times 10^{39}$\\
4494 & mekal2T,~~ $N_{\rm H}=5.13^{+630}_{-5.1}\times 10^{19} \rm{cm}^{-2}$, $kT_1=0.7^{+0.6}_{-0.4}$ keV, $kT_2=6^{+13}_{-2}$ keV & 13.2/14 & $1.6\times 10^{39}$\\
     & pow,~~ $N_{\rm H}=3^{+7}_{-3}\times 10^{20}$ cm$^{-2}$, $\Gamma=1.8\pm 0.3$ & 18.6/16 & $1.5\times 10^{39}$\\
4552 & pow,~~ $N_{\rm H}=6^{+6}_{-6}\times 10^{20}$ cm$^{-2}$, $\Gamma=2.0\pm 0.2$ & 43.7/43 & $3.7\times10^{39}$\\
     & compbb,~~ $N_{\rm H}=1.58\pm 0.04\times 10^{20}$ cm$^{-2}$, $kT=(9.06\pm 0.07)\times 10^{-2}$ keV, $\tau=7.03^{+0.01}_{-0.03}$ & 42.5/42 & $3.7\times10^{39}$\\
5055 & mekal2T,~~ $N_{\rm H}=0.5^{+0.4}_{-0.5}\times 10^{22}$ cm$^{-2}$, $kT_{1}=0.2^{+8}_{-0.2}$ keV, $kT_2=9^{+70}_{-2}$ keV & 4.3/10 & $8.3\times10^{38}$\\
\sidehead{\underline{\it Average Diffuse Emission Spectra}}
ellipticals & mekal+pow,~~ $N_{\rm H}=9^{+31}_{-8}\times 10^{19}~{\rm cm}^{-2}$, $kT=0.56^{+0.02}_{-0.03}$~keV, $\Gamma=1.3^{+0.1}_{-0.2}$ & 208.4/177 & $5.3\times 10^{38}$ \\
spirals     & mekal+pow,~~ $N_{\rm H}=(7\pm 3)\times 10^{20}$ cm$^{-2}$, $kT=0.57^{+0.02}_{-0.04}$ keV, $\Gamma=1.5^{+0.3}_{-0.1}$         & 237.1/236 & $8.5\times 10^{38}$ \\
\sidehead{\underline{\it Central Diffuse Emission Patches of Individual Galaxies}}
3628 & mekal2T,~~ $N_{\rm H}=6^{+5}_{-2}\times 10^{21}$ cm$^{-2}$, $kT_1=0.2^{+0.4}_{-0.1}$ keV, $kT_2=7^{+19}_{-4}$ keV & 18.07/18 & $4.6\times10^{38}$\\
     & pow,~~ $N_{\rm H}$=2$^{+2}_{-1}\times 10^{21}$ cm$^{-2}$, $\Gamma=1.5^{+0.3}_{-0.2}$ & 21.26/20 & $4.6\times10^{38}$\\
\sidehead{\underline{\it Brightest Off-Nuclear Point Sources of Galaxies Without a Central Source}}
3379 & pow,~~ $N_{\rm H}=1.1^{+0.7}_{-1.0}\times10^{21}$ cm$^{-2}$, $\Gamma=2.1_{-0.2}^{+0.3}$ & 31.8/27 & $1.3\times10^{39}$\\
     & mekal+pow,~~ $N_{\rm H}=1.8^{+1.8}_{-0.8}\times10^{21}$ cm$^{-2}$, $kT=0.08^{+0.14}_{-0.01}$ keV, $\Gamma=2.3^{+0.5}_{-0.3}$ & 27.2/25 & $1.5\times10^{39}$\\
\enddata
\tablenotetext{{\rm a}\;}{The two best-fitting models for sources with
more than 200 counts in their spectra. In the case of NGC~5055, only
one reasonable, well-fitting model was found. See \S\ref{sec:fits} for
explanation of models and their parameters.}
\end{deluxetable}

\begin{deluxetable}{cccccrrrrc}
\rotate
\tablenum{4}
\tablewidth{0in}
\tablecolumns{10}
\tablecaption{Diffuse Emssion Luminsity and Candidate AGN Properties}
\tablehead{
                                    & 
\multicolumn{2}{c}{Diffuse, 1 kpc Region\tablenotemark{a}} &
\multicolumn{2}{c}{Diffuse, 2\farcs 5 Region\tablenotemark{a}} &
\multicolumn{4}{c}{Candidate AGN\tablenotemark{a}}  \\
\noalign{\vskip -6pt}
\colhead{Galaxy}                    &
\multicolumn{2}{c}{\hrulefill}      &
\multicolumn{2}{c}{\hrulefill}      &
\multicolumn{4}{c}{\hrulefill}      \\
\colhead{(NGC)}                     &
\colhead{0.5--2 keV}                &
\colhead{2--10 keV}                 &
\colhead{0.5--2 keV}                &
\colhead{2--10 keV}                 &
\colhead{0.5--2 keV}                &
\colhead{2--10 keV}                 &
\colhead{$\REdd$\tablenotemark{b}}  &
\colhead{$\log R_{\rm X}$\tablenotemark{c}}          
}
\startdata

1553 & $3.3\times 10^{38}$ & $1.7\times 10^{38}$ & $1.6\times 10^{38}$ & $8.4\times 10^{37}$ &   $3.0\times 10^{39}$ &   $1.2\times 10^{40}$ & $  8.4\times 10^{-6}$ & $ -2.7$ \\
2681 & $1.3\times 10^{39}$ & $7.1\times 10^{38}$ & $9.9\times 10^{38}$ & $5.1\times 10^{38}$ &   $1.6\times 10^{38}$ &   $1.8\times 10^{38}$ & $  8.0\times 10^{-7}$ & $ -1.9$ \\
3379 & $4.3\times 10^{38}$ & $2.2\times 10^{38}$ & $1.1\times 10^{38}$ & $5.6\times 10^{37}$ &   $4.6\times 10^{37}$ &   $1.7\times 10^{37}$ & $  6.0\times 10^{-9}$ & $ -2.2$ \\
3507 & $7.8\times 10^{38}$ & $4.0\times 10^{38}$ & $6.4\times 10^{38}$ & $3.3\times 10^{38}$ & $< 2.5\times 10^{37}$ & $< 3.9\times 10^{37}$ & \dots                 & \dots   \\
3607 & $9.7\times 10^{37}$ & $5.0\times 10^{37}$ & $9.1\times 10^{37}$ & $4.7\times 10^{37}$ & $< 3.1\times 10^{37}$ & $< 5.0\times 10^{37}$ & $< 1.4\times 10^{-8}$ & \dots   \\
3608 & $8.7\times 10^{37}$ & $4.5\times 10^{37}$ & $8.3\times 10^{37}$ & $4.3\times 10^{37}$ &   $2.7\times 10^{38}$ &   $6.0\times 10^{38}$ & $  3.3\times 10^{-7}$ & $<-2.7$ \\ 
3628 & $1.1\times 10^{39}$ & $5.5\times 10^{38}$ & $8.9\times 10^{38}$ & $4.6\times 10^{38}$ & $< 3.1\times 10^{36}$ & $< 5.0\times 10^{36}$ & $< 4.4\times 10^{-9}$ & \dots   \\ 
4111 & $1.3\times 10^{39}$ & $7.0\times 10^{38}$ & $8.7\times 10^{38}$ & $4.5\times 10^{38}$ &   $7.0\times 10^{38}$ &   $3.7\times 10^{39}$ & $  4.1\times 10^{-6}$ & $ -3.0$ \\ 
4125 & $2.9\times 10^{38}$ & $1.5\times 10^{38}$ & $1.7\times 10^{38}$ & $9.2\times 10^{37}$ &   $3.4\times 10^{38}$ &   $5.4\times 10^{38}$ & $  1.5\times 10^{-7}$ & $<-2.7$ \\ 
4314 & $2.3\times 10^{39}$ & $1.2\times 10^{39}$ & $2.3\times 10^{38}$ & $1.2\times 10^{38}$ & $< 1.8\times 10^{37}$ & $< 2.9\times 10^{37}$ & $< 1.3\times 10^{-7}$ & \dots   \\ 
4374 & $2.5\times 10^{39}$ & $1.3\times 10^{39}$ & $1.3\times 10^{39}$ & $6.6\times 10^{38}$ &   $3.4\times 10^{39}$ &   $4.0\times 10^{39}$ & $  4.4\times 10^{-7}$ & $ -1.3$ \\ 
4438 & $3.7\times 10^{39}$ & $1.9\times 10^{39}$ & $3.9\times 10^{38}$ & $2.0\times 10^{38}$ &   $2.6\times 10^{39}$ &   $1.2\times 10^{39}$ & \dots                 &         \\ 
4457 & $6.2\times 10^{38}$ & $3.2\times 10^{38}$ & $4.2\times 10^{38}$ & $2.2\times 10^{38}$ &   $5.3\times 10^{38}$ &   $9.7\times 10^{38}$ & $  6.8\times 10^{-6}$ & $ -1.6$ \\ 
4494 & $7.6\times 10^{38}$ & $3.9\times 10^{38}$ & $7.2\times 10^{38}$ & $3.7\times 10^{38}$ &   $6.3\times 10^{38}$ &   $9.9\times 10^{38}$ & $  1.7\times 10^{-6}$ & $<-3.7$ \\ 
4552 & $1.7\times 10^{39}$ & $8.8\times 10^{38}$ & $1.2\times 10^{39}$ & $6.1\times 10^{38}$ &   $1.7\times 10^{39}$ &   $2.0\times 10^{39}$ & $  4.4\times 10^{-7}$ & $ -1.5$ \\ 
4636 & $7.7\times 10^{38}$ & $4.0\times 10^{38}$ & $5.4\times 10^{38}$ & $2.8\times 10^{38}$ & $< 3.9\times 10^{38}$ & $< 6.2\times 10^{38}$ & $< 2.7\times 10^{-7}$ & \dots   \\ 
5055 & $1.6\times 10^{39}$ & $8.3\times 10^{38}$ & $4.5\times 10^{38}$ & $2.3\times 10^{38}$ & $< 1.3\times 10^{38}$ & $< 2.0\times 10^{38}$ & $< 1.1\times 10^{-6}$ & \dots   \\ 
5866 & $2.1\times 10^{38}$ & $1.1\times 10^{38}$ & $5.4\times 10^{37}$ & $2.8\times 10^{37}$ &   $1.0\times 10^{37}$ &   $3.1\times 10^{38}$ & $  4.3\times 10^{-7}$ & $ -1.5$ \\ 
7331 & $6.6\times 10^{38}$ & $3.4\times 10^{38}$ & $4.2\times 10^{38}$ & $2.2\times 10^{38}$ & $< 2.1\times 10^{37}$ & $< 3.4\times 10^{37}$ & $< 6.0\times 10^{-8}$ & \dots   \\ 

\enddata

\tablenotetext{a\;}{Luminosities are given in erg~s$^{-1}$.}

\tablenotetext{b\;}{The Eddington ratio; see equation~(\ref{eq:eddington}) in
\S\ref{sec:discussion}.}

\tablenotetext{c\;}{The X-ray radio-loudness parameter; see
equation~(\ref{eq:Rx}) in \S\ref{sec:individual}.}

\end{deluxetable}

\normalsize

\clearpage

\begin{figure}
{\large For a complete preprint, including images, see
{\tt http://www.astro.psu.edu/users/mce/preprints/}}
\bigskip
\caption{Montage of the {\it Chandra} images of the target galaxies. For
each galaxy the image encompass 1~kpc in radius centered on the center
of the galaxy. The size of each pixel is {0\farcs}5. For each target,
the left image is for the energy band 0.5-2~keV, the middle image is
for the energy band 2-10~keV and the right image is the hardness ratio
map (see \S\ref{sec:analysis}). The grey scale represents the number
of counts per pixel and is logarithmic and adjusted to give the
maximal dynamical range for each image. North is up and east is to the
left.\label{fig:images}}
\end{figure}

\clearpage 

\begin{figure}
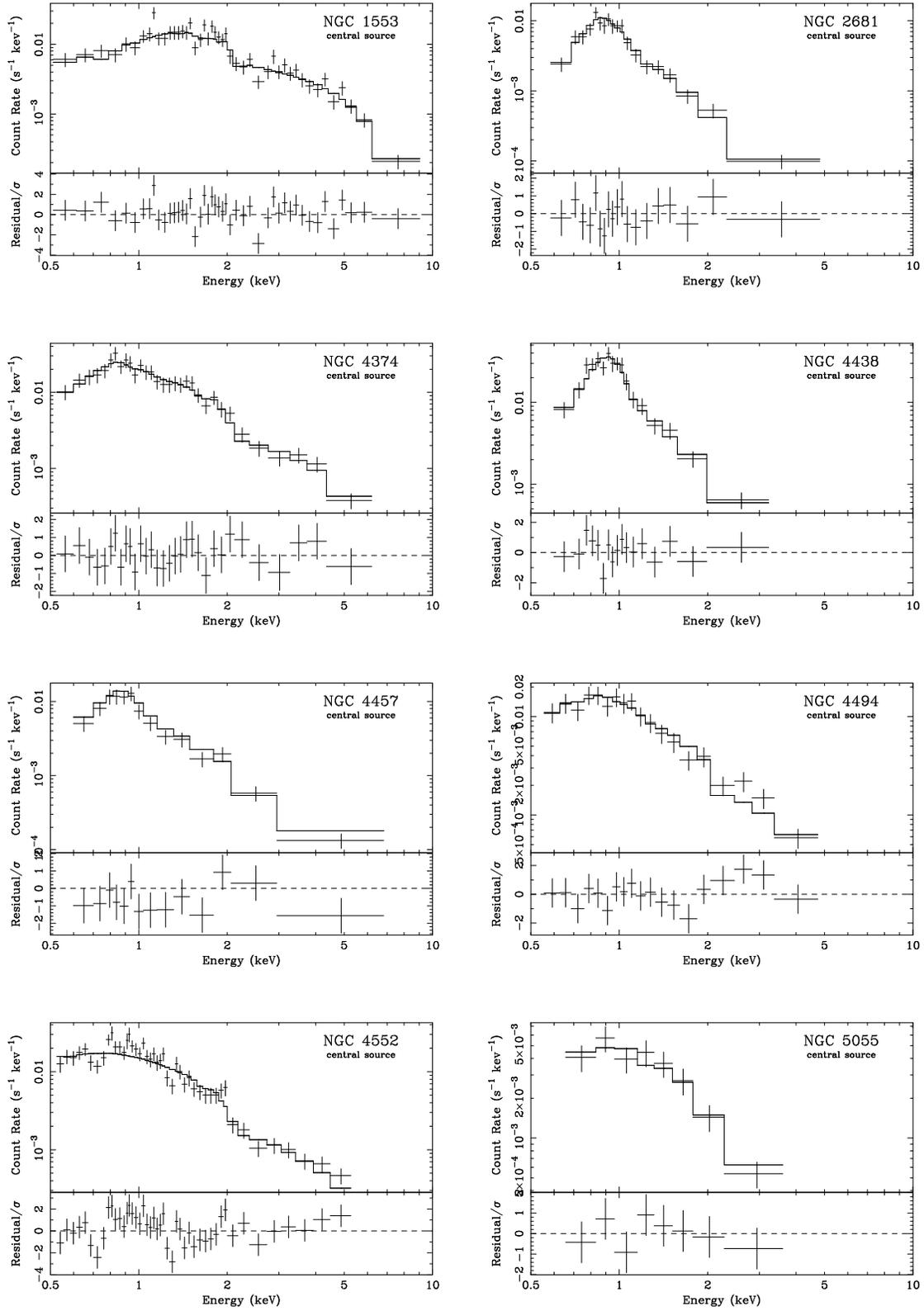

\centerline{
\rotatebox{-90}{\resizebox{!}{3in}{\includegraphics{f2a.eps}}}
\rotatebox{-90}{\resizebox{!}{3in}{\includegraphics{f2b.eps}}}
}
\medskip
\centerline{
\rotatebox{-90}{\resizebox{!}{3in}{\includegraphics{f2c.eps}}}
\rotatebox{-90}{\resizebox{!}{3in}{\includegraphics{f2d.eps}}}
}
\medskip
\centerline{
\rotatebox{-90}{\resizebox{!}{3in}{\includegraphics{f2e.eps}}}
\rotatebox{-90}{\resizebox{!}{3in}{\includegraphics{f2f.eps}}}
}
\medskip
\centerline{
\rotatebox{-90}{\resizebox{!}{3in}{\includegraphics{f2g.eps}}}
\rotatebox{-90}{\resizebox{!}{3in}{\includegraphics{f2h.eps}}}
}
\medskip
\caption{Spectra of the sources listed in Table~3 with the
best-fitting model superposed. The lower frame shows the residuals 
scaled by the error bars. The models and their parameters are given
in Table~3.\label{fig:spectra}}
\end{figure}

\begin{figure}
\centerline{
\rotatebox{-90}{\resizebox{!}{3in}{\includegraphics{f2i.eps}}}
\rotatebox{-90}{\resizebox{!}{3in}{\includegraphics{f2j.eps}}}
}
\medskip
\centerline{
\rotatebox{-90}{\resizebox{!}{3in}{\includegraphics{f2i.eps}}}
\rotatebox{-90}{\resizebox{!}{3in}{\includegraphics{f2j.eps}}}
}
\medskip
\centerline{Figure~\ref{fig:spectra} -- {\it continued}.}
\end{figure}

\begin{figure}
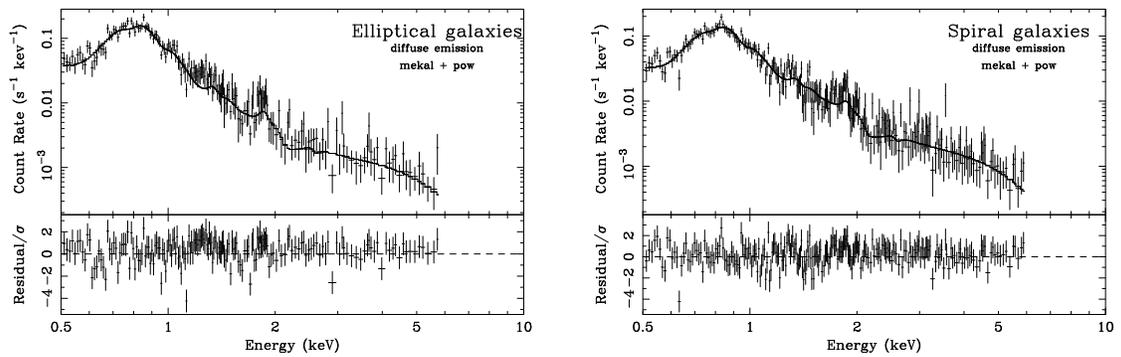

\centerline{
\rotatebox{-90}{\resizebox{!}{3in}{\includegraphics{f3a.eps}}}
\rotatebox{-90}{\resizebox{!}{3in}{\includegraphics{f3b.eps}}}
}
\caption{Stacked diffuse emission spectra for the elliptical galaxies
(left panel) and the spiral galaxies( right panel). Details of the
models fitted to the data are given in {\S}6 of the
text.\label{fig:stackedspec}}
\end{figure}

\begin{figure}
\centerline{
\resizebox{!}{2.5in}{\includegraphics{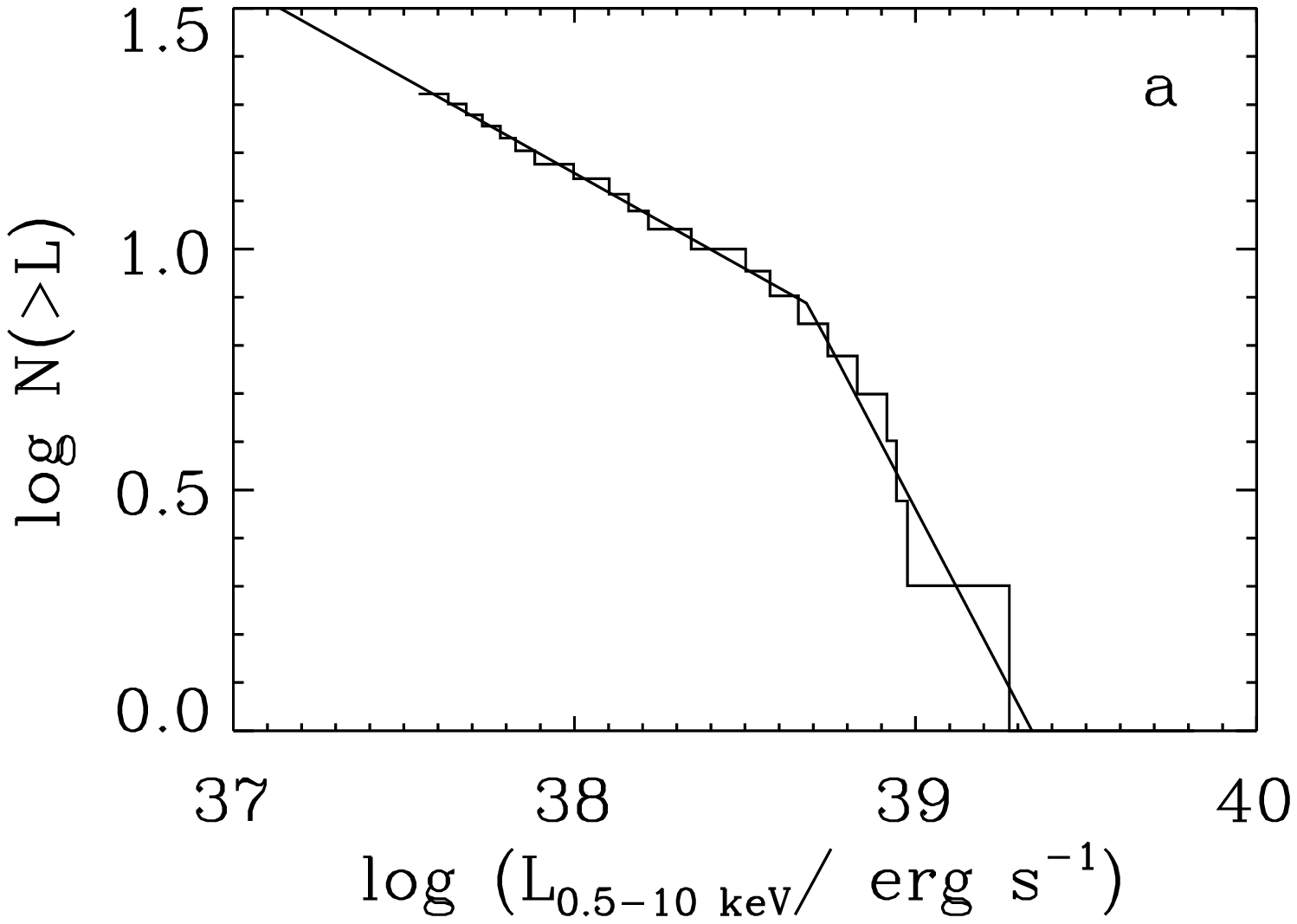}}
\resizebox{!}{2.5in}{\includegraphics{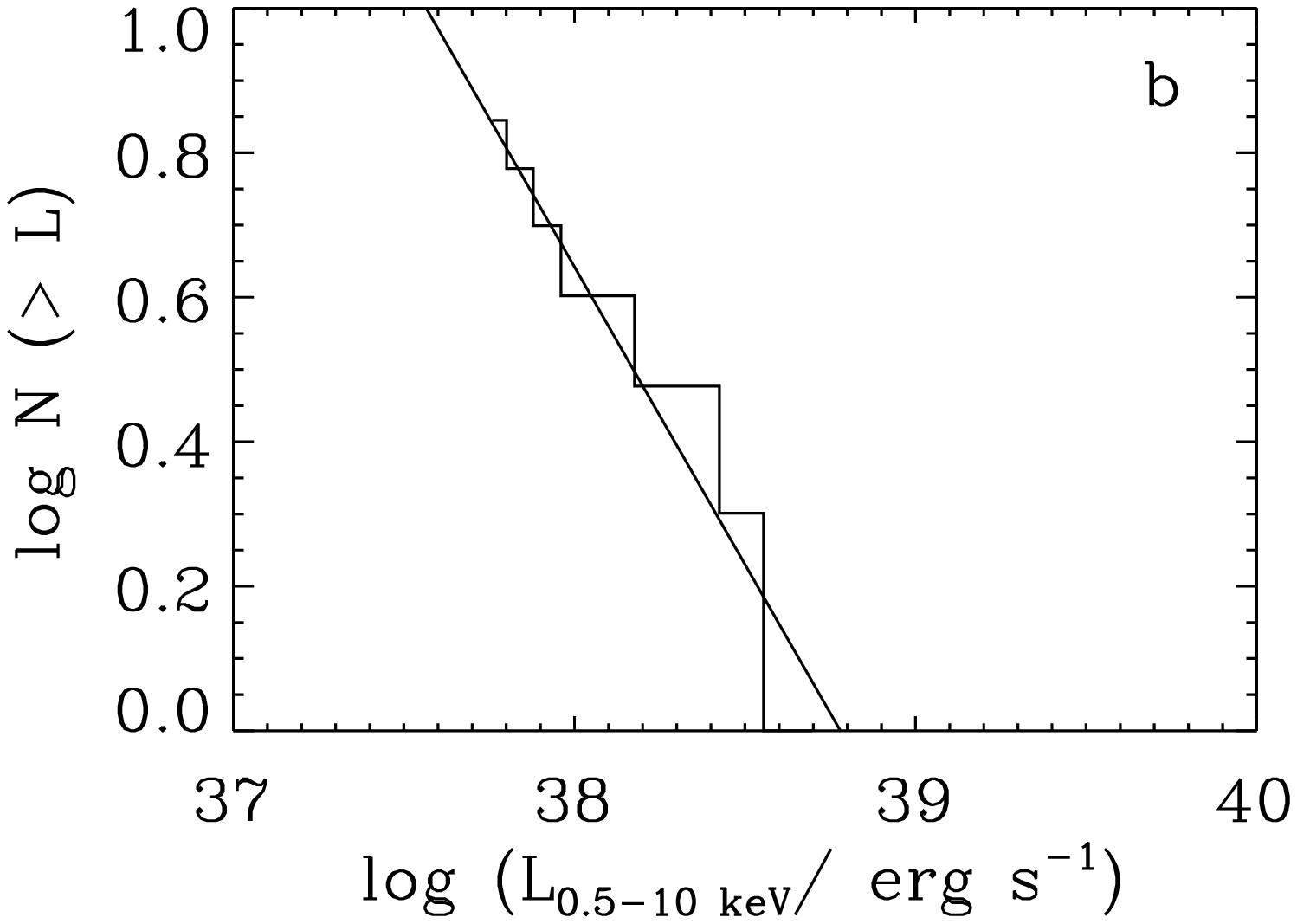}}
}
\caption{Histograms showing the cumulative luminosity functions for
the point sources in all the elliptical (a) and all the spiral
galaxies (b).  The continuous lines show single or broken power-law
fits to the data (see {\S}5 of the text for full details).
\label{fig:lumfunc}}
\end{figure}

\begin{figure}
\vskip -0.5truein
\centerline{\resizebox{!}{3.5in}{\includegraphics{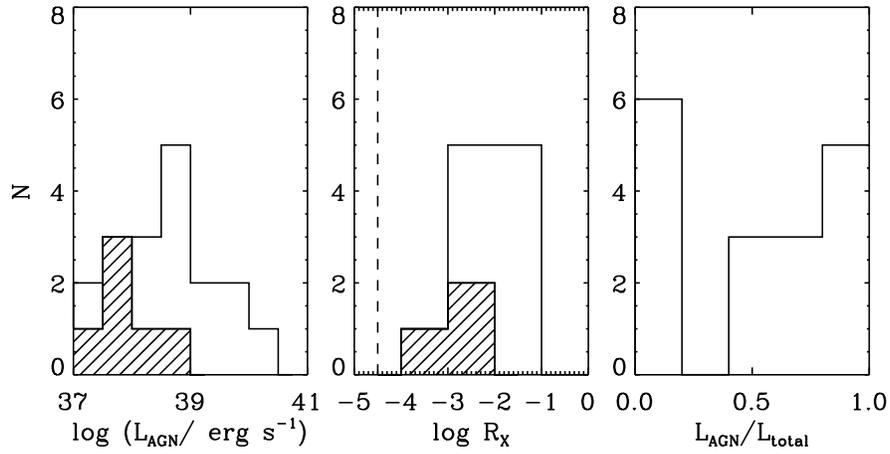}}}
\caption{Histograms showing the distribution of $L^{\rm AGN}_{\rm
2-10\; keV}$, log $R_X$, and $L^{\rm AGN}_{\rm
0.5-10\;keV}/L^{2{\farcs}5}_{\rm 0.5-10\; keV}$ for the galaxies of
our sample. The hatched regions represent upper limits. The threshold
between radio-loud and radio-quiet sources is shown on the log $R_X$
histogram as a dashed line.}
\end{figure}

\begin{figure}
\centerline{
\resizebox{!}{2.5in}{\includegraphics{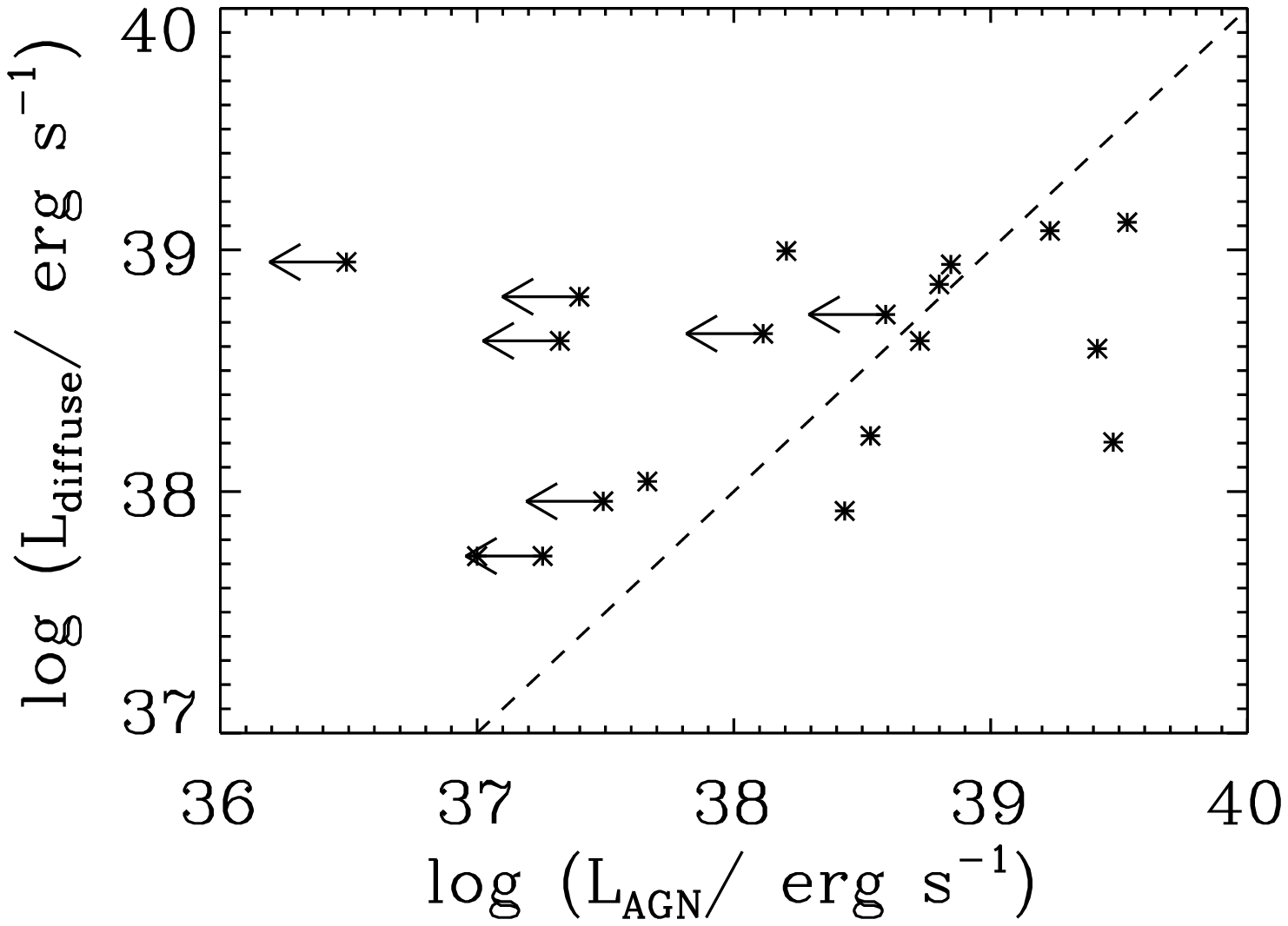}}
\resizebox{!}{2.5in}{\includegraphics{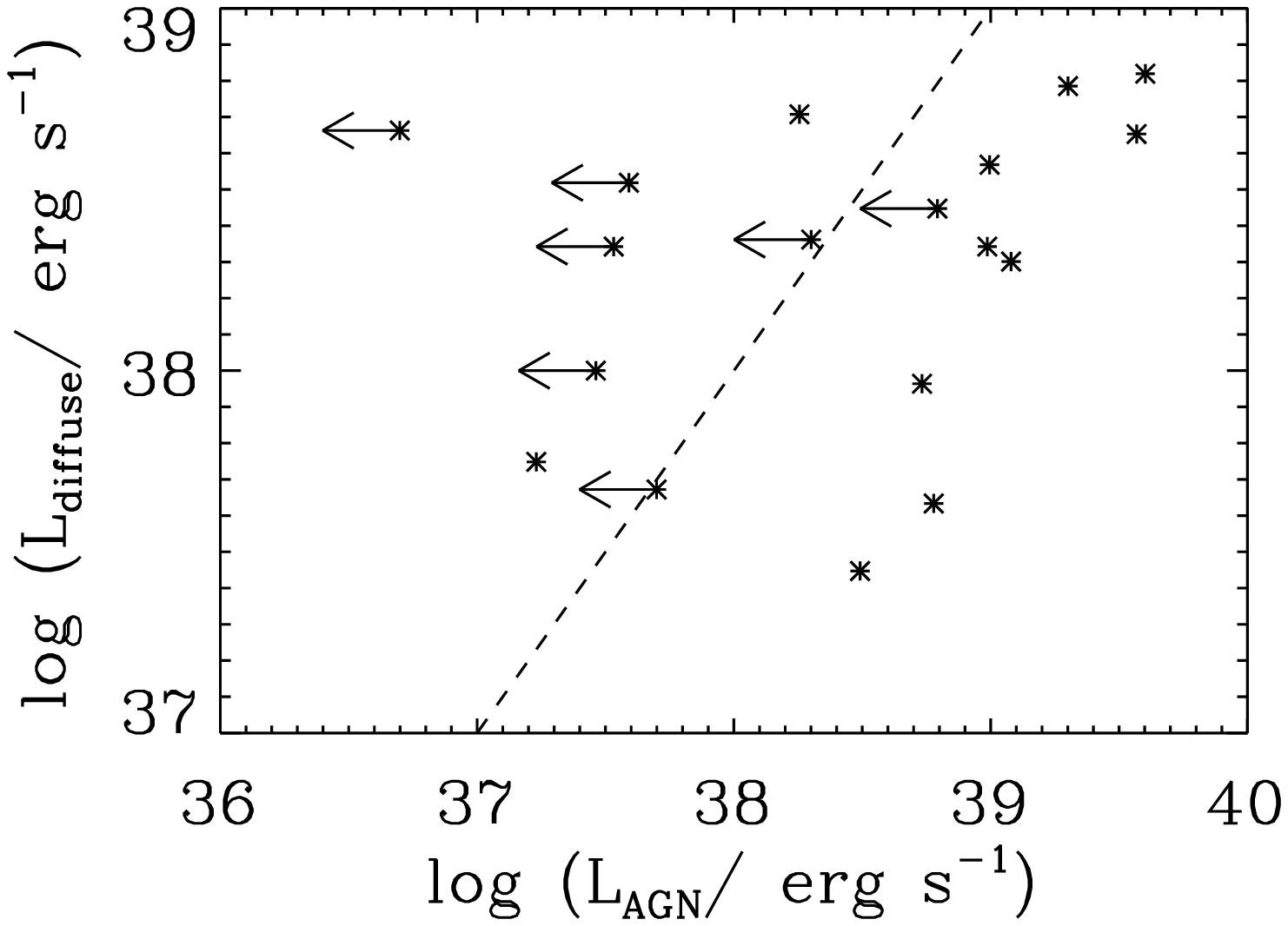}}
}
\caption{Luminosity from the diffuse emission in the inner {2\farcs}5
as a function of the luminosity from the AGN in the 0.5--2 keV band
(top) and 2--10 keV band (bottom). The dashed line represents equal
luminosity from the diffuse emission equals and from the
AGN. \label{fig:balance}}
\end{figure}

\begin{figure}
\centerline{\resizebox{!}{3.5in}{\includegraphics{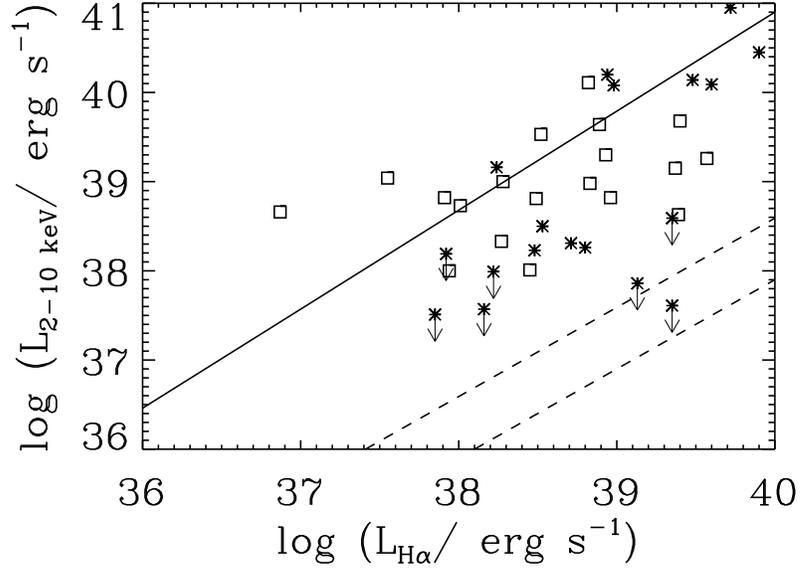}}}
\caption{$L_{\rm 2-10\; keV}-L_{H\alpha}$ plot for our sample (squares)
and the objects from the Ho et~al. (2001) sample (stars). The X-ray
luminosity is measured within the inner {2\farcs}5 region.  The solid
line represents the best fit found by Ho et~al. (2001) for
low-redshift quasars and Seyfert~1 galaxies. The dotted lines
represent the range of luminosities where a starburst would
fall.\label{fig:hoandmine}}
\end{figure}

\begin{figure}
\centerline{
\resizebox{!}{2.5in}{\includegraphics{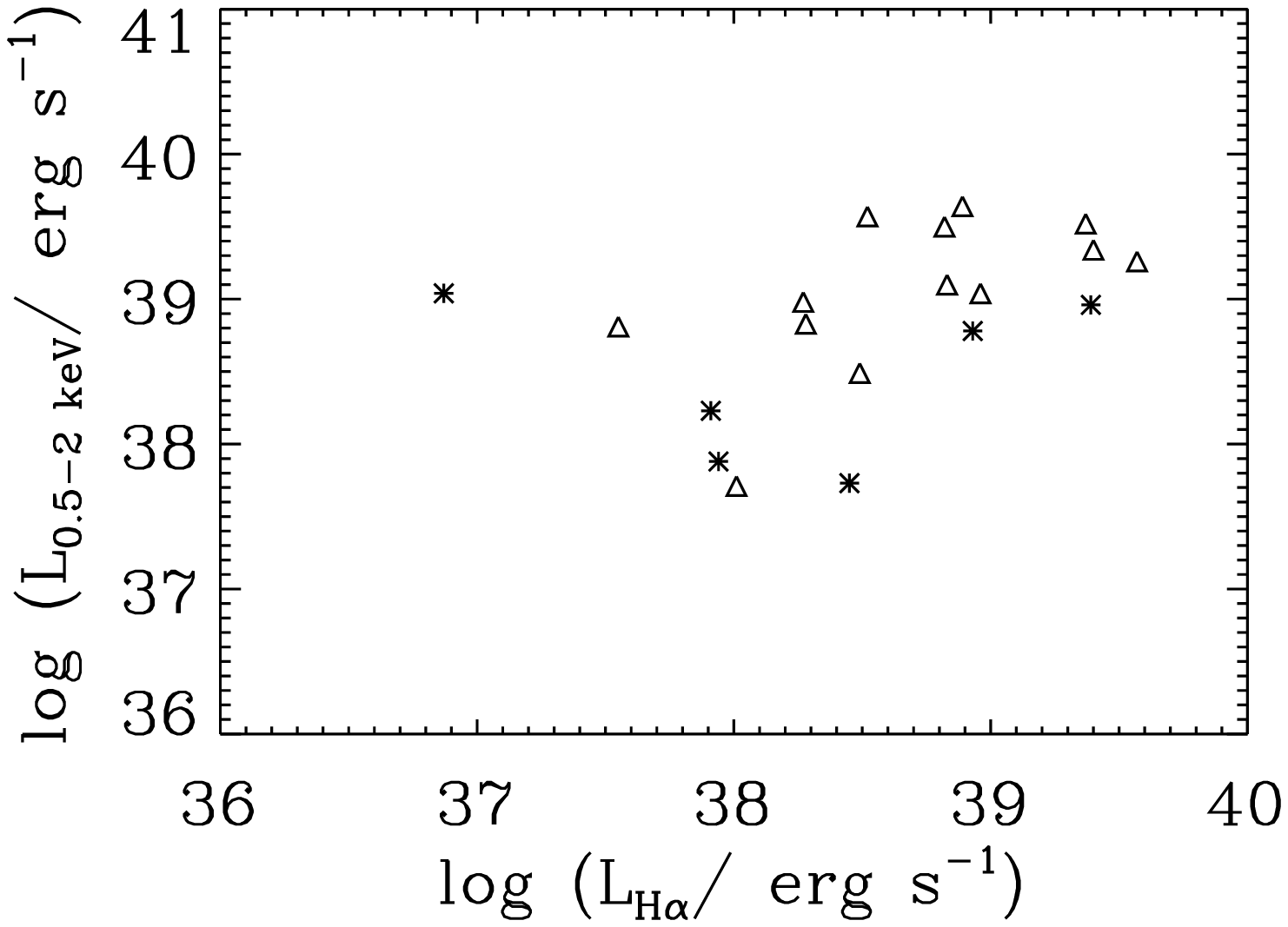}}
\resizebox{!}{2.5in}{\includegraphics{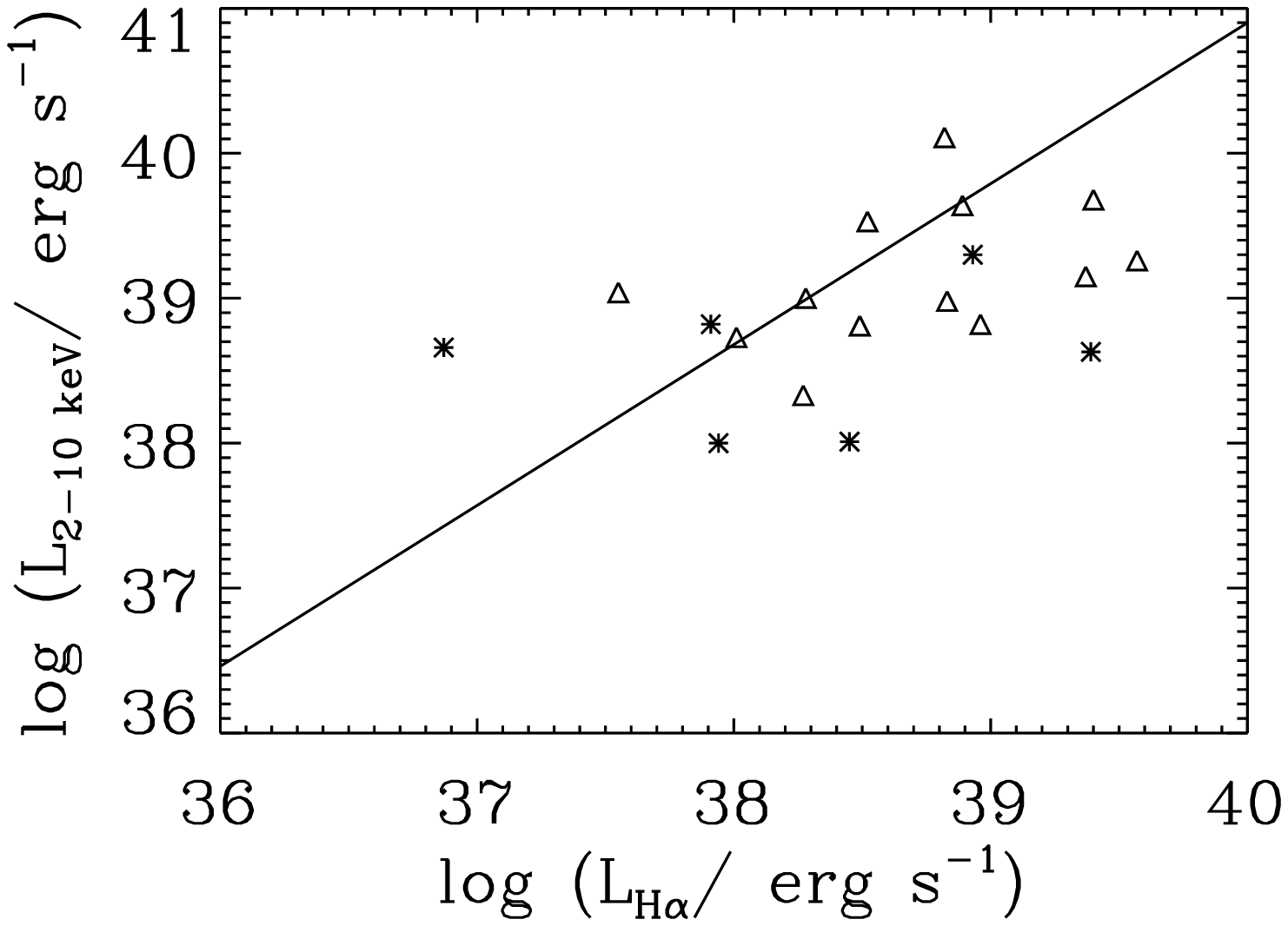}}
}
\caption{$L_{\rm 0.5-2\; keV}-L_{H\alpha}$ and $L_{\rm2-10\;
keV}-L_{H\alpha}$ diagrams. The stars represent LINERs whose X-ray
luminosity of the central 2\farcs 5 region is dominated by diffuse
emission while triangles represent AGN-dominated LINERs. The solid
line shows the extrapolation of the correlation found for low-redshift
quasars and Seyfert 1 galaxies (from Ho et
al. 2001).\label{fig:lxvslha}}
\end{figure}

\end{document}